\UseRawInputEncoding
\documentclass[sn-mathphys-num]{sn-jnl}

\usepackage{graphicx}%
\usepackage{multirow}%
\usepackage{amsmath,amssymb,amsfonts}%
\usepackage{amsthm}%
\usepackage{mathrsfs}%
\usepackage[title]{appendix}%
\usepackage{xcolor}%
\usepackage{textcomp}%
\usepackage{manyfoot}%
\usepackage{booktabs}%
\usepackage{algorithm}%
\usepackage{algorithmicx}%
\usepackage{algpseudocode}%
\usepackage{listings}%

\theoremstyle{thmstyleone}%
\theoremstyle{thmstyletwo}%

\theoremstyle{thmstylethree}%
\newtheorem{definition}{Definition}%

\begin{document}

\title[Article Title]{A new perspective on brain stimulation interventions: Optimal stochastic tracking control of brain network dynamics}


\author*[1]{\fnm{Kangli} \sur{Dong}}\email{kanglidong@stu.edu.cn}
\equalcont{These authors contributed equally to this work.}

\author[2]{\fnm{Siya} \sur{Chen}}\email{siyachen4-c@my.cityu.edu.hk}
\equalcont{These authors contributed equally to this work.}

\author[3]{\fnm{Ying} \sur{Dan}}\email{yingdan@link.cuhk.edu.hk}

\author[4]{\fnm{Lu} \sur{Zhang}}\email{3413037@zju.edu.cn}

\author[1]{\fnm{Xinyi} \sur{Li}}\email{24xyli1@stu.edu.cn}

\author[1]{\fnm{Wei} \sur{Liang}}\email{21wliang@stu.edu.cn}

\author[5]{\fnm{Yue} \sur{Zhao}}\email{24520190154819@stu.xmu.edu.cn}

\author[6]{\fnm{Yu} \sur{Sun}}\email{yusun@zju.edu.cn}

\affil*[1]{\orgdiv{Department of Biomedical Engineering, College of Engineering}, \orgname{Shantou University}, \orgaddress{\city{Shantou}, \postcode{515063}, \state{Guangdong}, \country{China}}}

\affil[2]{\orgdiv{Department of Computer Science}, \orgname{City University of Hong Kong}, \orgaddress{\city{Hong Kong}, \postcode{999077}, \state{Hong Kong}, \country{China}}}

\affil[3]{\orgdiv{Department of biomedical engineering}, \orgname{Chinese University of Hong Kong}, \orgaddress{\city{Hong Kong}, \postcode{999077}, \state{Hong Kong}, \country{China}}}

\affil[4]{\orgdiv{Department of Rehabilitation, Sir Run Run Shaw Hospital}, \orgname{Zhejiang University School of Medicine, Zhejiang University}, \orgaddress{\city{Hangzhou}, \postcode{310027}, \state{Zhejiang}, \country{China}}}

\affil[5]{\orgdiv{Department of Urology, Xiang’an Hospital of Xiamen University}, \orgname{Xiamen University}, \orgaddress{\city{Xiamen}, \postcode{361102}, \state{Fujian}, \country{China}}}

\affil[6]{\orgdiv{Key Laboratory for Biomedical Engineering of Ministry of Education of China}, \orgname{Zhejiang University}, \orgaddress{\city{Hangzhou}, \postcode{310007}, \state{Zhejiang}, \country{China}}}


\abstract{Network control theory (NCT) has recently been utilized in neuroscience to facilitate our understanding of brain stimulation effects and explore optimal paradigms. A particularly useful branch of NCT is optimal control, which focuses on applying theoretical and computational principles of control theory to design optimal strategies to achieve specific goals in neural processes.
However, most existing research focuses on optimally controlling brain network dynamics from the original state to a target state at a specific time point. Additionally, these studies overlook the influence of neuronal noise on dynamics, thereby simplifying the network dynamics to a linear deterministic system.
In this paper, we present the first investigation of considering stochastic brain network dynamic system and introducing optimal stochastic tracking control strategy to synchronize the dynamics of the brain network to a target dynamics rather than to a target state at a specific time point.
To accomplish this, our analysis utilized fMRI data from healthy groups, and cases of stroke and post-stroke aphasia. For all participants, we utilized a gradient descent optimization method to estimate the parameters (e.g., the coupled matrix and the variance matrix) for the brain network dynamic system.
We then utilized optimal stochastic tracking control techniques to drive original unhealthy dynamics by controlling a certain number of nodes to synchronize with target healthy dynamics.
Results show that the energy associated with optimal stochastic tracking control is negatively correlated with the intrinsic average controllability of the brain network system, while the energy of the optimal state approaching control is significantly related to the target state value. 
Additionally, for a 100-dimensional brain network system, controlling the five nodes with the lowest tracking energy can achieve relatively acceptable dynamics control effects. 
Our results suggest that stochastic tracking control is more aligned with the objective of brain stimulation interventions, and is closely related to the intrinsic characteristics of the brain network system, potentially representing a new direction for future brain network optimal control research.}




\maketitle

\section{Introduction}\label{sec1}

The brain, functions as a vast and intricate network, with various brain parcellation templates categorizing it into distinct regions based on either anatomical or functional structures. The functional connectivity (FC) between these regions is often modeled as nodes and edges in a graph, creating a FC matrix that illustrates the network of interactions between brain regions. With the increasing availability of extensive public datasets and open-access, integrated toolboxes \cite{rubinov2010complex, whitfield2012conn, wang2015gretna}, the application of graph theory to network neuroscience has become increasingly prevalent \cite{baggio2014functional, kim2017abnormal, dennis2014functional, deletoile2017graph, ashtiani2018altered, farahani2019application, agosta2014disrupted, tao2020different, chen2021disrupted}. By representing neural connections in the brain as graphs, network topological analysis has become a widely used method for investigating fundamental characteristics of brain networks, such as the small-world property, which is marked by efficient, short-range connections between nodes \cite{bassett2006small, gallos2012conundrum}, the presence of hub nodes with a power-law distribution of node degrees \cite{zamora2010cortical, guye2010graph,dong2024meso}, the modular organization of brain regions into functional modules or subsystems \cite{bullmore2009complex, sporns2014contributions}, and rich club structures characterized by tightly interconnected hub nodes \cite{van2011rich}. 

\begin{figure}[tbp]
\centering
\includegraphics[scale=0.8]{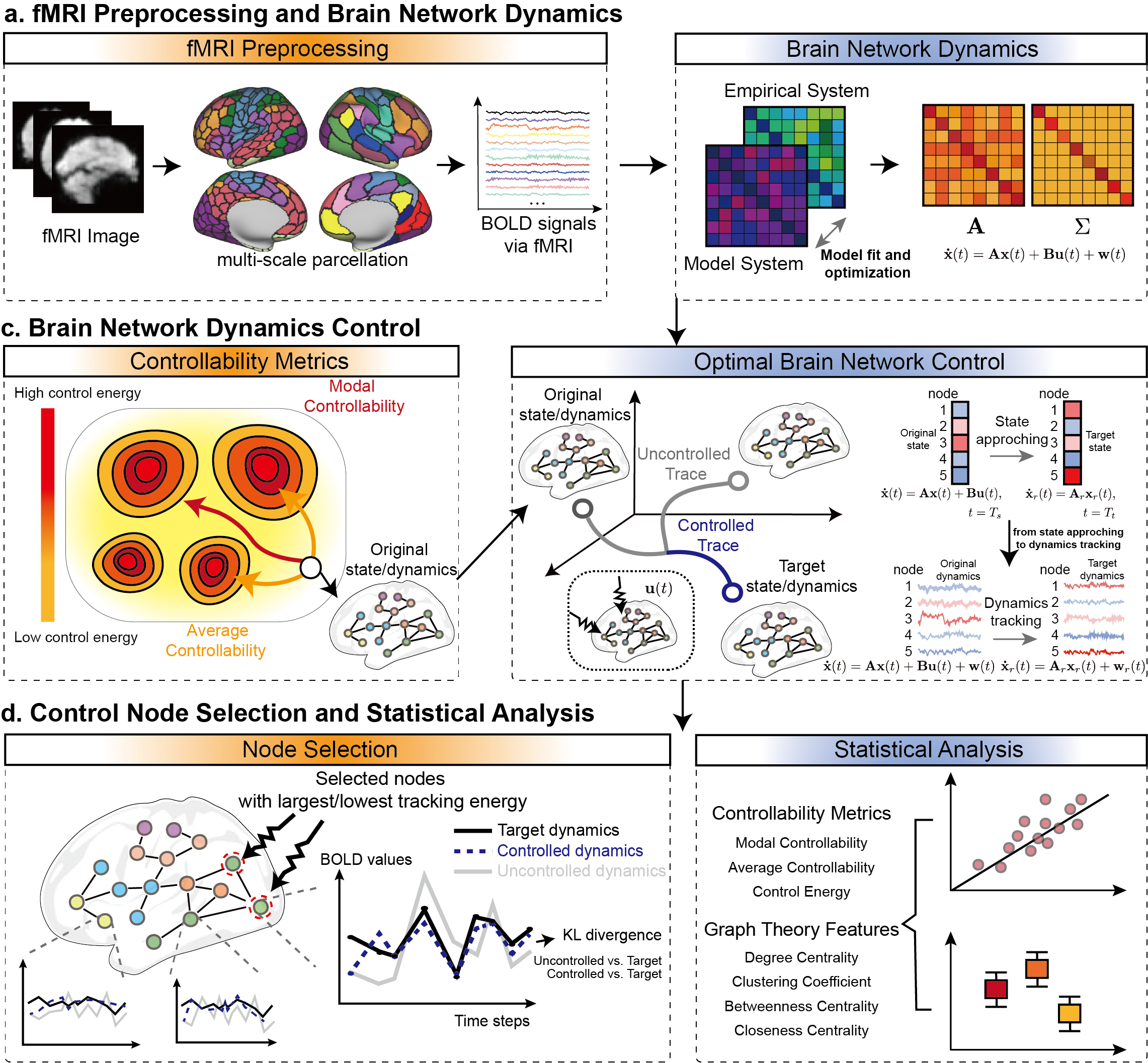}
\caption{Framework. \textbf{a.} fMRI Data Preprocessing and Brain Network Model Construction. The preprocessing phase includes fMRI denoising and obtaining the BOLD signal after cortical parcellation. Here we utilized the 100-ROI and 400-ROI templates from Schaefer2018 parcellation \cite{schaefer2018local}, at different scales. In the brain network model construction phase, we employed the gradient descent optimization method to estimate the $\mathbf{A}$ and $\mathbf{\Sigma}$ of the network system, thereby making the network model closely approximate the fMRI dynamics. The coupled matrix $\mathbf{A}$ and variance matrix $\mathbf{\Sigma}$ were further used to carry out the optimal stochastic tracking control and optimal state approaching control. \textbf{b.} Brain Network Control. First, we estimated the two intrinsic properties of brain network systems, namely average controllability and modal controllability, to explore the relationship between these properties and the optimal control characteristics. Second, as a prerequisite for optimal state approaching control and optimal stochastic tracking control, we extracted the original brain states and dynamics (i.e., time series) for each stroke and post-stroke aphasia patient, and extracted target brain states and dynamics from the healthy group as control targets. Finally, we executed two different optimal control strategies and examined the characteristics of their primary performance (i.e., control energy). \textbf{c.} Control Node Selection and Statistical Analysis. We conducted statistical analyses on the control energy, the controllability of the brain network, and the nodal metrics under two optimal control strategies. Additionally, based on the control energy (i.e., the tracking energy of optimal stochastic tracking control), we attempted to explore the control effects using KL divergence with different numbers of control nodes by incrementally adding control nodes.}\label{fig1}
\end{figure}

In addition to understanding the topological changes in brain networks under pathological conditions, brain stimulation interventions have emerged as a potential effective therapeutic approach. 
In recent years, the clinical use of transcranial magnetic stimulation (TMS) \cite{hallett2007transcranial} and transcranial direct current stimulation (tDCS) \cite{feng2013review} has shown promising results, driven by advancements in non-invasive brain stimulation \cite{polania2018studying}. 
Theoretically, Network Control Theory (NCT) offers us the opportunity to understand how these stimulation effect the brain dynamics and theoretically explore optimal stimulation paradigms. NCT posits that the brain operates as a dynamic system, transitioning through various activation patterns, or brain states, over time \cite{greene2023everyone}. NCT quantifies brain network properties based on regional interactions and overall network dynamics, offering a computational framework to elucidate how functional brain dynamics arise from underlying network structures  \cite{karrer2020practical}. In the context of NCT, a brain network is deemed controllable if it can be controlled/pushed into different active `states', defined as coordinated neurophysiological activities across brain regions at specific moments \cite{greene2023everyone}. Based on NCT, the optimal brain network control is particularly influential. In essence, it generates time-varying inputs that drive the brain network system to a target state. These inputs are optimized to minimize the squared magnitude of the input signal integrated over time, known as the `control energy', and as the distance from a target state, typically set to be equal to the target state or an $N \times 1$ zero vector for a $N$-dimension system. Optimal brain network control has been carried out to study executive function and brain network development \cite{cornblath2020temporal,parkes2022asymmetric,sun2023network}, epilepsy \cite{stiso2019white,cooper2021mapping}, schizophrenia \cite{braun2019brain}, psychiatric disorders \cite{parkes2021network,mahadevan2023alprazolam}, and psychedelic research \cite{singleton2022receptor,singleton2023time}, etc. 

Existing optimal brain network control methods are predominantly based on the assumption of simplest linear systems ($\mathbf{\dot{x}}(t)=\mathbf{A} \mathbf{x}(t)+\mathbf{B} \mathbf{u}(t)$). This framework is reasonable, as brain dynamics can be approximated by linear dynamics to some extent \cite{honey2009predicting}. However, the current assumption still has overlooked two significant factors that cannot be ignored: First, the current assumption do not consider the spontaneous dynamics in brain networks. The current dynamical equations do not include a noise term, and the assumption cannot satisfy the condition that brain dynamics passively evolve according to their own dynamic inertia when uncontrolled.
Second, the essence of dynamics is time series, and the assumption currently in use defines a state as a discrete state at a specific time point within the dynamics. A single discrete state cannot capture the characteristics of brain dynamics, and merely pushing one state to another is insufficient.

In this investigation, we incorporate the aforementioned factors into optimal brain network control, and present the first investigation of introducing the concept of \textit{optimal stochastic tracking control} to reconsider the brain network control problem (see the framework at Fig.~\ref{fig1}). The goal of optimal stochastic tracking control is to drive original dynamics towards target dynamics, making their statistical properties similar, which is more aligned with the objective of achieving brain dynamic control (right panel of Fig.~\ref{fig1}.\textbf{b}). Additionally, to implement optimal stochastic tracking control in practical functional magenetic resonance imaging (fMRI) measurements, we introduced and improved a gradient descent optimization method to obtain the statistical variance and diagonally symmetric connectivity matrix of the brain network system (right panel of Fig.~\ref{fig1}.\textbf{a}). Using fMRI data from stroke and post-stroke aphasia cases as examples, we attempted to drive the disease dynamics towards healthy dynamics. 
Stroke is a cerebral disorder resulting from damage to cerebral blood vessels due to various factors, leading to either localized or widespread brain injury \cite{feigin2021global}. As a major cause of aphasia, stroke can impair language abilities, with patients losing their capacity to communicate due to acute brain damage \cite{bernhardt2017agreed, varkanitsa2024insights}. The recovery process is complicated by both short-term and long-term speech impairments, as well as the diverse transitions observed across different forms of post-stroke aphasia \cite{sheppard2021diagnosing}. Individuals experiencing chronic aphasia (persisting for over six months) may attain substantial improvements through speech therapy \cite{wade1986aphasia}.
By using stroke and post-stroke aphasia data and comparing the results with existing optimal control strategies (referred to as \textit{optimal state approaching control} in subsequent sections), we validated that the proposed new framework can reflect the intrinsic controllability characteristics of the brain network system. We also explored the minimum cost required to achieve an acceptable control effect, aiming to provide basis for clinical brain stimulation paradigms.

\begin{figure}[bp]
\centering
\includegraphics[scale=0.35]{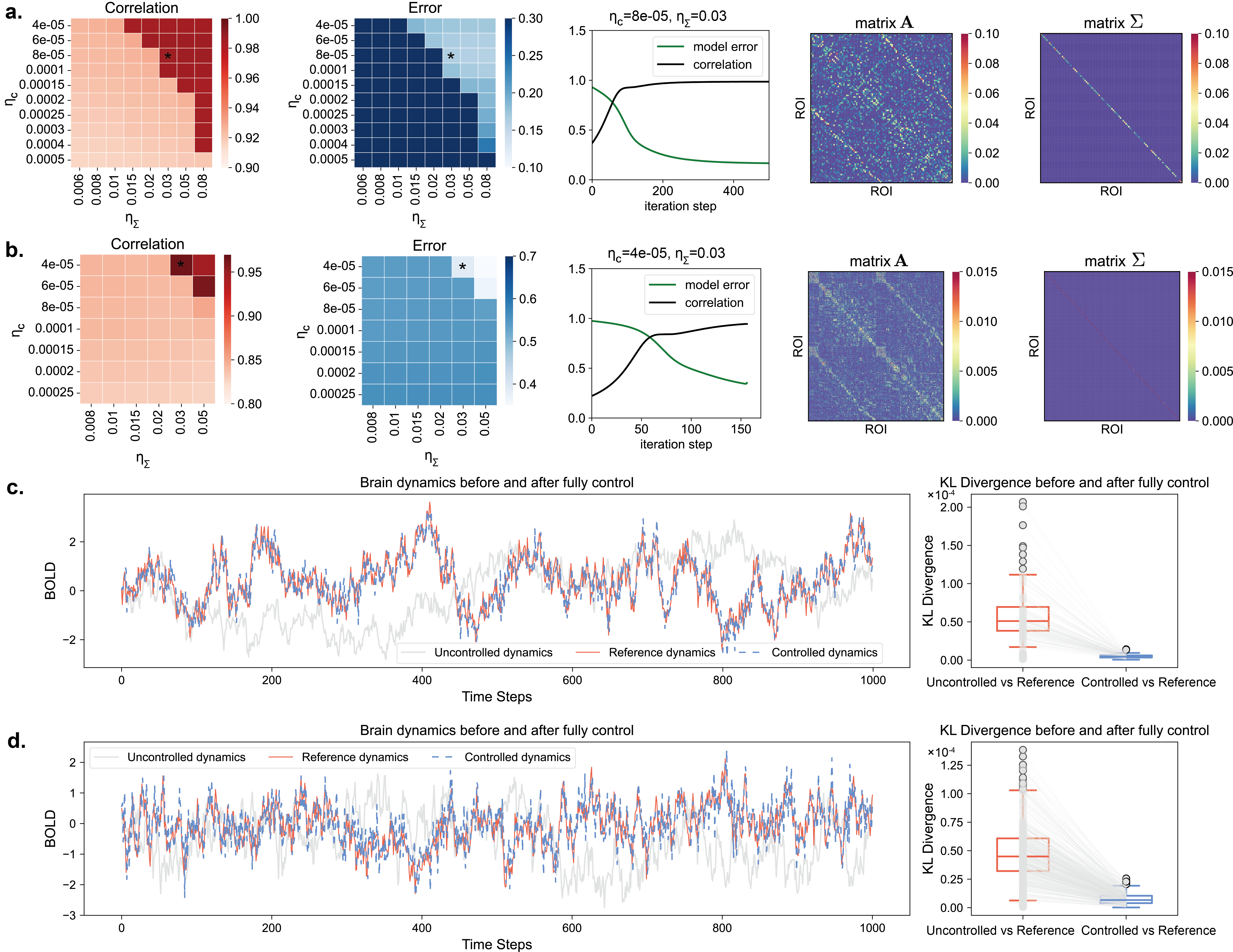}
\caption{The brain network system and the dynamics after fully control (all nodes are controlled using optimal stochastic tracking control) for a representative participant (ID: M2121), along with the control effects. \textbf{a.} Based on the 100-ROI parcellation, from left to right: the correlation between the optimized brain network system and the target dynamics under different parameter pairs, the error with respect to the target dynamics ($\star$ indicates the selected optimal parameter pair), the optimization process under the optimal parameter pair, and the optimized $\mathbf{A}$ matrix and $\mathbf{\Sigma}$ matrix. \textbf{b.} Based on the 400-ROI parcellation, the content is the same as in part \textbf{a}. \textbf{c.} Based on the 100-ROI parcellation, the target dynamics, the controlled dynamics, and the uncontrolled dynamics (ROI = 1) are visualized as time series. Additionally, a comparison is made between the KL divergence of the controlled dynamics versus the target dynamics, and the KL divergence of the uncontrolled dynamics versus the target dynamics for each node in the brain network system. \textbf{d.} Based on the 400-ROI parcellation, the content is the same as in part \textbf{c}. }\label{fig2}
\end{figure}

\section{Results}\label{sec2}
In this investigation, we present the first investigation of introducing optimal stochastic tracking control framework to brain network control, aiming to provide insights for optimal brain stimulation interventions. Using stroke and post-stroke aphasia as examples, in this section, we first provide a case study of a participant to demonstrate the effectiveness of the control strategy. We then present the spatial distribution characteristics of tracking energy and its relationship with the intrinsic controllability of brain systems. Further, we attempt to select control nodes based on tracking energy to investigate how the dynamics of the brain network system can be controlled to the minimum extent. Finally, we provide the results of optimal state approaching control for comparison. We applied a 100-ROI and a 400-ROI templates from Schaefer2018 parcellation \cite{schaefer2018local} for simulation. Due to space constraints, part of results for the 100-ROI brain network system and the correlation analysis of nodal metrics are available in the supplementary materials.

\subsection{Fitting model for brain network dynamic system}
To better control brain network dynamics rather than performing discrete state approaching, we first obtained the optimal parameters (coupled matrix $\mathbf{A}\in R^{N \times N}$ and variance matrix $\mathbf{\Sigma}\in R^{N \times N}$) for each participant's brain network system using gradient descent optimization (as shown in Fig.~\ref{fig2}.\textbf{a} and \textbf{b}, using the participant with ID 2121 as an example). Since the optimization process involves two important parameters ($\eta_\mathbf{C}$ and $\eta_\mathbf{\Sigma}$) that need to be adjusted, we targeted two objectives - correlation and error (the average correlation and distance between the zero-lag and one-lag covariance matrices of the optimization model and the actual dynamics), and searched the optimal parameter pairs through a simple grid search. These pairs needed to simultaneously satisfy sufficiently high correlation and sufficiently low error. As shown in Fig.~\ref{fig2}.\textbf{a}, the optimal parameter pair for M2121's 100-ROI brain network system is $\eta_\mathbf{C} = 8e-05$ and $\eta_\mathbf{\Sigma} = 0.03$, while for the 400-ROI system, the optimal parameter pair is $\eta_\mathbf{C} = 4e-05$ and $\eta_\mathbf{\Sigma} = 0.03$ (Fig.~\ref{fig2}.\textbf{b}). After selecting the optimal parameter pair, we can observe the corresponding optimization processes for the 100-ROI and 400-ROI brain networks (the middle panels in Fig.~\ref{fig2}.\textbf{a} and \textbf{b}). Finally, we obtained the optimized $\mathbf{A}$ and $\mathbf{\Sigma}$ (the two right panels in Fig.~\ref{fig2}.\textbf{a} and \textbf{b}). This computational process was applied to all patients.

\subsection{Fully control effect using optimal stochastic tracking control}
After obtaining the optimal brain system parameters, we performed fully optimal control (controlling all nodes) for all stroke and aphasia participants. The control effects are shown in Fig.~\ref{fig2}.\textbf{c} and \textbf{d}, which visualize the dynamics (ROI = 1) and statistical results of Kullback-Leibler (KL) divergence of each node/ROI in the 100-ROI and 400-ROI brain network systems, respectively. It can be observed that the controlled dynamics is very close to the target dynamics, with the KL divergence between them approaching zero. In contrast, the uncontrolled dynamics is significantly different from the target dynamics. The KL divergence estimates indicate that the statistical similarity of the uncontrolled dynamics to the target dynamics for each node is larger than that of the controlled dynamics, demonstrating that the tracking control achieved excellent results.

\begin{figure}[htbp]
\centering
\includegraphics[scale=0.2]{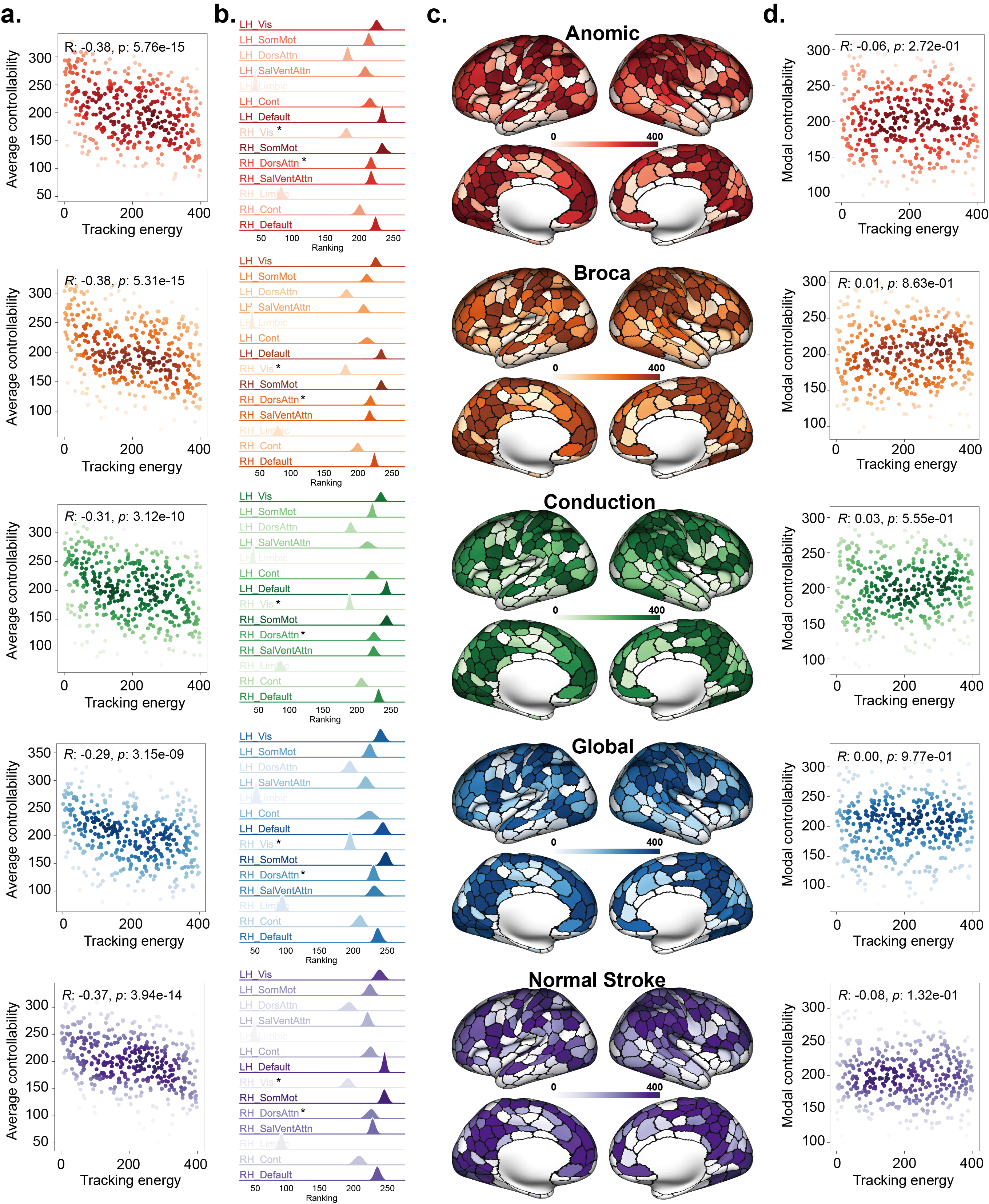}
\caption{The tracking energy ranking (from the largest to the lowest) of the 400-ROI brain network system, from top to bottom: Anomic, Broca, Conduction, Global, Normal Stroke. \textbf{a.} The correlation between the ranking of tracking energy and the ranking of average controllability, with $\textit{R}$ values and $\textit{p}$ values provided. \textbf{b.} The distribution of the averaged ranking of tracking energy among brain network nodes in each resting-state system. $\star$ indicates significant differences between the five groups from top to bottom (using the Kruskal–Wallis test and FDR correction). \textbf{c.} Visualization of the ranking of tracking energy on the cortex. \textbf{d.} The correlation between the ranking of tracking energy and the ranking of modal controllability, with $\textit{R}$ values and $\textit{p}$ values provided.}\label{fig3}
\end{figure}

\subsection{Spatial distribution of tracking energy and its relationship with network controllability}
We further investigated the distribution of tracking energy across the cortex under fully control and quantified its relationship with the intrinsic controllability of brain network. From Fig.~\ref{fig3}.\textbf{a} and \textbf{d}, it can be observed that for the 400-ROI brain network system of all five different groups of participants, there is a significant negative correlation between the ranking of tracking energy and the average controllability of the brain network system (Anomic: $R = -0.38, p = 5.76e-15$, Broca: $R = -0.38, p = 5.31e-15$, Conduction: $R = -0.31, p = 3.12e-10$, Global: $R = -0.29, p = 3.15e-09$, Normal stroke: $R = -0.37, p = 3.94e-14$). However, there is no significant correlation between the ranking of tracking energy and modal controllability (Anomic: $R = -0.06, p = 2.72e-01$, Broca: $R = 0.01, p = 8.63e-01$, Conduction: $R = 0.03, p = 5.55e-01$, Global: $R = 0.00, p = 9.77e-01$, Normal stroke: $R = -0.08, p = 1.32e-01$). For the 100-ROI brain network system, we found highly consistent results (see Supplementary Fig. S1), where the ranking of tracking energy also showed a significant negative correlation with average controllability (Anomic: $R = -0.47, p = 6.31e-07$, Broca: $R = -0.50, p = 1.43e-07$, Conduction: $R = -0.47, p = 8.93e-07$, Global: $R = -0.42, p = 1.19e-05$, Normal stroke: $R = -0.51, p = 4.77e-08$) and no significant correlation with modal controllability (Anomic: $R = 0.07, p = 5.18e-01$, Broca: $R = 0.15, p = 1.42e-01$, Conduction: $R = 0.10, p = 3.03e-01$, Global: $R = 0.14, p = 1.52e-01$, Normal stroke: $R = 0.07, p = 9.87e-01$).

Additionally, we observed spatially heterogeneous distribution in the ranking of tracking energy across different resting-state systems (Fig.~\ref{fig3}.\textbf{b} and \textbf{c}). It can be seen that the left and right Limbic regions have the highest rankings, indicating that they consume the highest tracking energy for optimal control. In contrast, the left Vis, Default, and right SomMot regions have the lowest rankings, indicating that they consume the lowest tracking energy. Further, we found that when comparing the five different groups, there were significant differences ($p \textless 0.05$) in the ranking of tracking energy for the right Vis and right DorsAttn regions (using the Kruskal–Wallis test and FDR correction), as detailed in the supplementary materials (Supplementary Fig. S2). For the 100-ROI brain network system, significant differences ($p \textless 0.05$) were observed only in the right Vis region (Supplementary Figs. S1 and S2). Additionally, we examined the relationship between tracking energy and the topological metrics of network nodes (Supplementary Figs. S3 and S4), and found that for the 400-ROI brain network system, tracking energy is positively correlated with degree and negatively correlated with clustering coefficient.

\begin{figure}[htbp]
\centering
\includegraphics[scale=0.17]{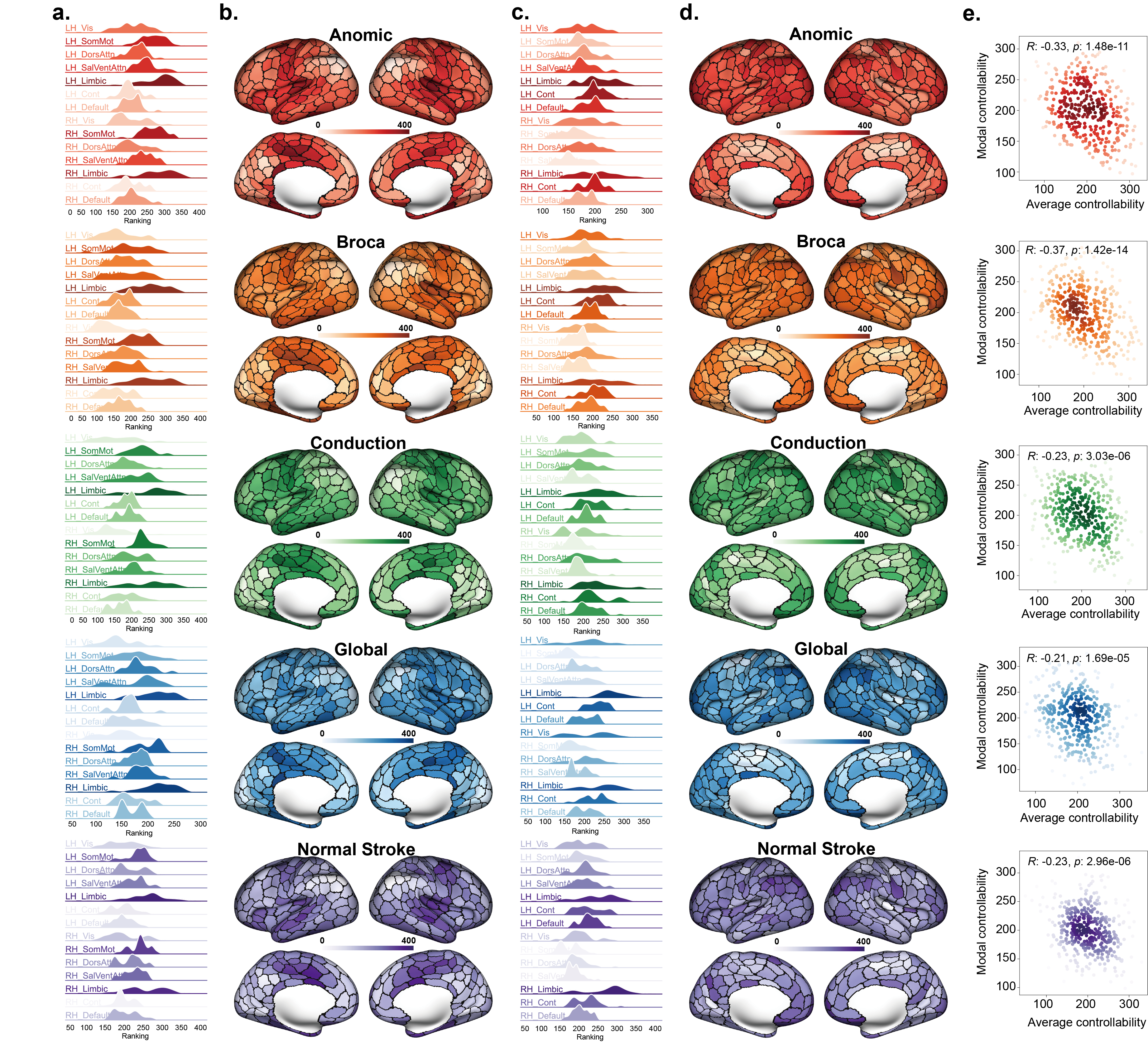}
\caption{The average controllability and modal controllability ranking (from the largest to the lowest) of the 400-ROI brain network system, from top to bottom: Anomic, Broca, Conduction, Global, Normal Stroke. \textbf{a.} The distribution of the averaged ranking of average controllability among brain network nodes in each resting-state system. \textbf{b.} Visualization of the ranking of average controllability on the cortex. \textbf{c.} The distribution of the averaged ranking of modal controllability among brain network nodes in each resting-state system. \textbf{d.} Visualization of the ranking of modal controllability on the cortex. \textbf{e.} The correlation between the ranking of average controllability and the ranking of modal controllability, with $\textit{R}$ values and $\textit{p}$ values provided.}\label{fig4}
\end{figure}

\subsection{Characteristics of average controllability and modal controllability}
Similarly, we quantified the correlation relationship between the two types of intrinsic controllability and explored their spatial distributions. From Fig.~\ref{fig4}.\textbf{a} and \textbf{b}, the spatial heterogeneity in the ranking of average controllability across different resting-state systems is not as apparent as the distinct distribution of tracking energy. However, consistent with the significant negative correlation observed between tracking energy and average controllability, the left and right Limbic regions have the lowest rankings for average controllability, and with no significant distributional differences among the five groups. From Fig.~\ref{fig4}.\textbf{c} and \textbf{d}, it can be seen that for the 400-ROI brain network system, the spatial distribution of modal controllability is more uniform across the cortex, with no significant distributional differences among the five groups. Consistent with previous studies \cite{gu2015controllability,gu2017optimal,wilmskoetter2022language}, there is a significant negative correlation between average controllability and modal controllability for all five groups (as shown in Fig. ~\ref{fig4}.\textbf{e}, Anomic: $R = -0.33, p = 1.48e-11$, Broca: $R = -0.37, p = 1.42e-14$, Conduction: $R = -0.23, p = 3.03e-06$, Global: $R = -0.21, p = 1.69e-05$, Normal stroke: $R = -0.23, p = 2.96e-06$). For the 100-ROI brain network system, the results are completely consistent with those of the 400-ROI system, with a significant negative correlation between average controllability and modal controllability (as shown in the Supplementary Fig. S3, Anomic: $R = -0.45, p = 3.10e-06$, Broca: $R = -0.55, p = 4.36e-09$, Conduction: $R = -0.40, p = 8.76e-05$, Global: $R = -0.38, p = 1.09e-04$, Normal stroke: $R = -0.29, p = 3.09e-03$). 

Following previous studies  \cite{gu2015controllability,gu2017optimal}, we examined the relationship between average/modal controllability and the topological metrics of nodes (Supplementary Figs. S6-S9). Consistent with previous studies \cite{gu2015controllability,gu2017optimal}, we found that for the 100-ROI brain network system, average controllability is positively correlated with degree in Anomic, Broca, and Conduction groups. However, for the 400-ROI brain network system, average controllability is negatively correlated with degree and positively correlated with clustering coefficient. This discrepancy may be related to the presence of numerous negative values in the 400-ROI system, as unlike structural and functional brain networks, the optimized brain network system allows the occurrence of negative connectivities. For modal controllability, consistent with previous studies \cite{gu2015controllability,gu2017optimal}, significantly negative correlation can be observed between modal controllability and degree, and significantly positive correlation can be observed between modal controllability and clustering coefficient, in which the significance is more pronounced in the 400-ROI system.

\begin{figure}[htbp]
\centering
\includegraphics[scale=0.5]{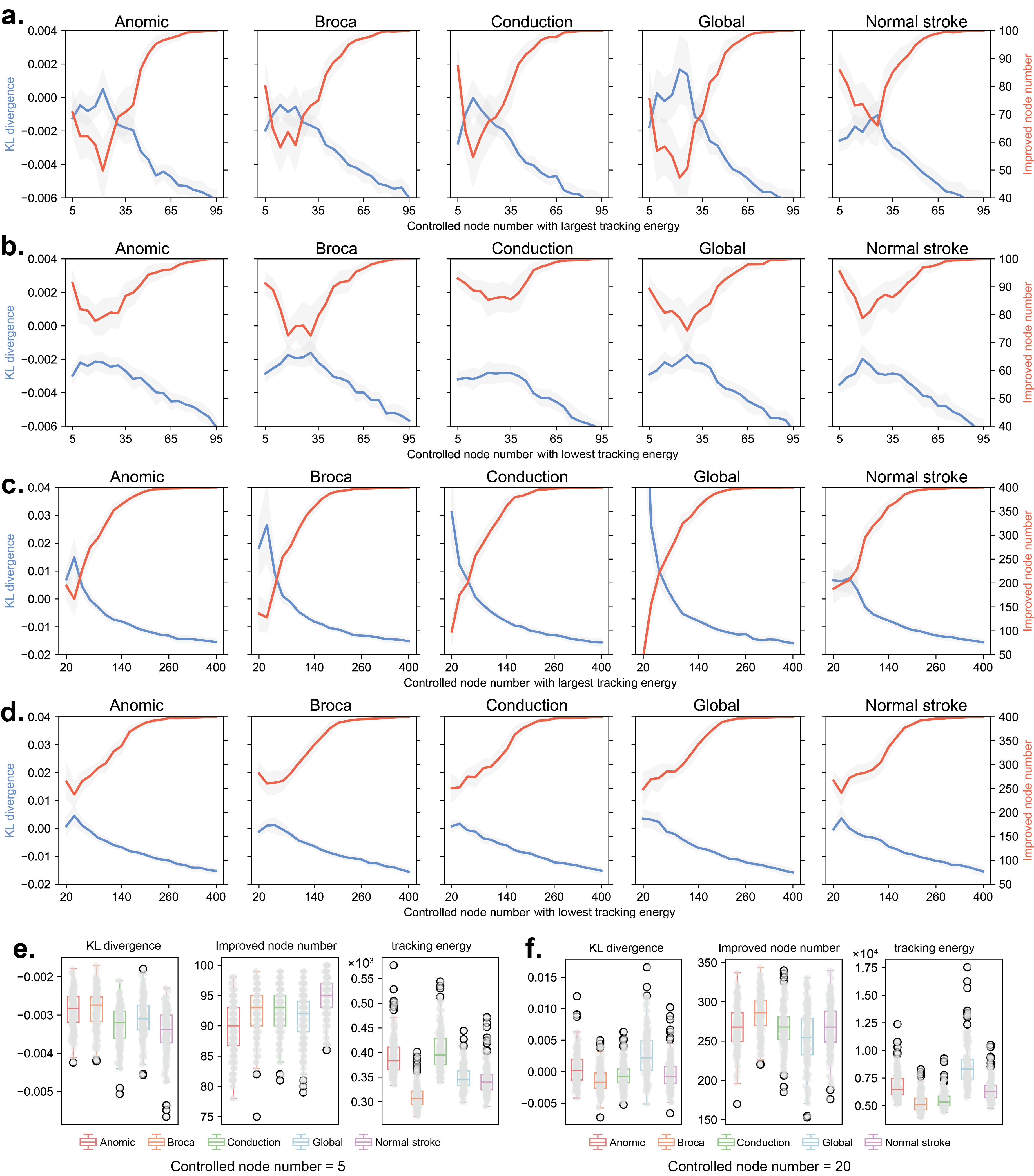}
\caption{The control effects with different numbers of control nodes, from left to right: Anomic, Broca, Conduction, Global, Normal Stroke.
\textbf{a.} For the 100-ROI brain network system, the total KL divergence of the system and the number of nodes with improved KL divergence when control nodes are incrementally added in descending order of tracking energy. Controlled nodes are added in increments of 5.
\textbf{b.} For the 100-ROI brain network system, the total KL divergence of the system and the number of nodes with improved KL divergence when control nodes are incrementally added in ascending order of tracking energy. Controlled nodes are added in increments of 5.
\textbf{c.} Based on the 400-ROI parcellation, the content is the same as in part \textbf{a}. Controlled nodes are added in increments of 20.
\textbf{d.} Based on the 400-ROI parcellation, the content is the same as in part \textbf{b}. Controlled nodes are added in increments of 20.
\textbf{e.} The control effect for the 100-ROI brain network system when the controlled nodes are the 5 nodes with the lowest tracking energy.
\textbf{f.} The control effect for the 400-ROI brain network system when the controlled nodes are the 20 nodes with the lowest tracking energy.}\label{fig5}
\end{figure}

\subsection{Relationship between number of controlled nodes and tracking control effect}
To facilitate clinical translation, we attempted to select controlled nodes based on tracking energy to investigate how the brain network system can achieve a basically controllable state at the lowest possible cost. We employed two different strategies for both the 100-ROI and 400-ROI brain network systems: selecting nodes with the largest tracking energy and selecting nodes with the lowest tracking energy. We gradually increased the number of nodes to observe changes in the total KL divergence between the controlled dynamics and the target dynamics, as well as the number of improved nodes (nodes with reduced KL divergence after control compared to before control). We constructed an averaged model for each group, and considering the significant influence of noise on stochastic tracking control process, we performed 200 simulations for each averaged model to obtain final results.

Fig.~\ref{fig5}.\textbf{a} and \textbf{b} show the changes in KL divergence and the number of improved nodes for the 100-ROI brain network system under the largest and lowest tracking energy strategies, respectively. The shaded gray areas represent the confidence intervals from 200 simulations. It can be observed that when the number of controlled nodes is relatively low, selecting nodes with lowest tracking energy achieves better control effects than selecting nodes with largest tracking energy. For the five different groups, when the number of controlled nodes is 5, selecting lowest tracking energy nodes can control more than 90\% of the system's nodes (Fig.~\ref{fig5}.\textbf{b}), whereas selecting largest tracking energy nodes can only control 70-90\% of the system's nodes. Additionally, within the range of 5-35 nodes, a higher number of nodes does not necessarily result in better control effects.

For the 400-ROI brain network system, the results are similar to those of the 100-ROI system. However, the increased dimensionality of the system modeling makes the system more difficult to control. Under the strategy of selecting lowest tracking energy nodes, using 20 nodes can only control 60-75\% of the system's nodes (250-300 nodes).

We further compared the control effects of the five groups in the 100-ROI brain network system with 5 controlled nodes and the 400-ROI brain network system with 20 controlled nodes. The results show that when the system is modeled in 100 dimensions, the normal stroke group is easier to control, with more than 90\% of nodes experiencing a decrease in KL divergence (Fig.~\ref{fig5}.\textbf{e}). When the system is modeled in 400 dimensions, number of effective controllable nodes is usually below 300 (Fig.~\ref{fig5}.\textbf{f}). Additionally, the tracking energy differ between the 100-ROI and 400-ROI systems. For the 100-ROI system, Anomic and Conduction aphasia require the highest energy, whereas for the 400-ROI system, Global aphasia requires the highest energy (Fig.~\ref{fig5}.\textbf{e} and \textbf{f}). This discrepancy arises partly from the modeling dimensions and partly from the potential asymmetry of controlled nodes.

\begin{figure}[bp]
\centering
\includegraphics[scale=0.55]{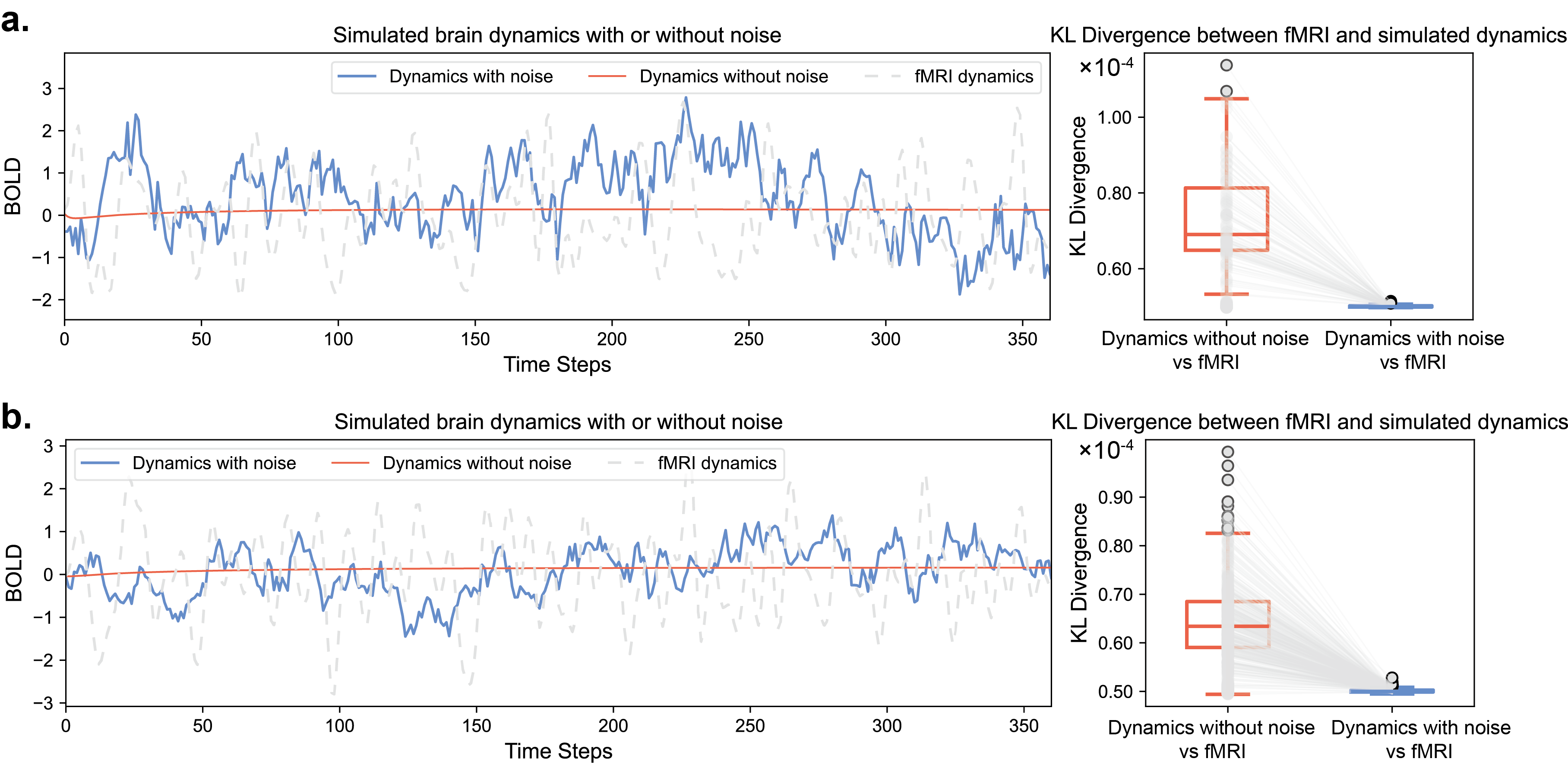}
\caption{The simulated brain dynamics with noise ($\mathbf{\dot{x}}(t)=\mathbf{A} \mathbf{x}(t) + \mathbf{w}(t)$) and without noise ($\mathbf{\dot{x}}(t)=\mathbf{A} \mathbf{x}(t)$) for a representative participant (ID: M2121). \textbf{a.} Based on the 100-ROI parcellation, the fMRI dynamics, the simulated dynamics with noise, and the simulated dynamics without noise (ROI = 1) are visualized as time series. Additionally, a comparison is made between the KL divergence of the simulated dynamics with noise versus the fMRI dynamics, and the KL divergence of the simulated dynamics without noise versus the fMRI dynamics for each node in the brain network system. \textbf{b.} Based on the 400-ROI parcellation, the content is the same as in part \textbf{a}. }\label{fig6}
\end{figure}

\begin{figure}[tp]
\centering
\includegraphics[scale=0.35]{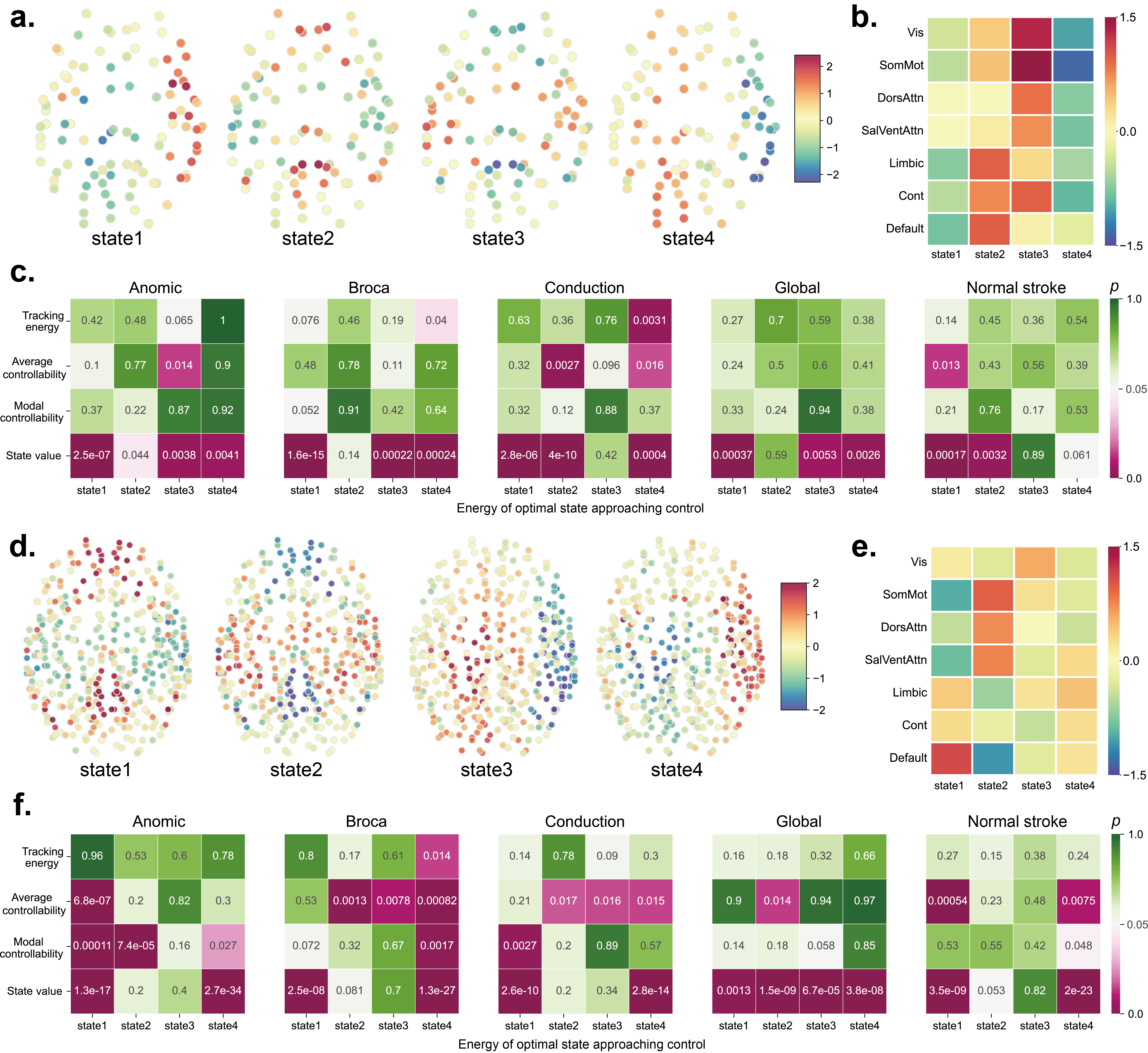}
\caption{Results of optimal state approaching control energy. \textbf{a.} The four target states of the 100-ROI brain network system obtained through clustering. \textbf{b.} The distribution of activation features of the four target states of the 100-ROI brain network system in each resting-state system. \textbf{c.} For the 100-ROI brain network system, the p-values of the correlations between the optimal state approaching control energy for the four target states and tracking energy, average controllability, modal controllability, and the states themselves. Significance thresholds are defined by 0.05, with values above 0.05 shown in green and values below 0.05 shown in red. \textbf{d.} Based on the 400-ROI parcellation, the content is the same as in part \textbf{a}. \textbf{e.} Based on the 400-ROI parcellation, the content is the same as in part \textbf{b}. \textbf{f.} Based on the 400-ROI parcellation, the content is the same as in part \textbf{c}.}\label{fig7}
\end{figure}

\subsection{Comparison with related work}
\subsubsection{Simulated brain dynamics with and without neuronal noise}
In this section, we first demonstrate the brain dynamics considering neuronal noise compared to the network model without considering noise, as illustrated in Fig.~\ref{fig6}. We use the data of a representative participant with ID 2121 as an example, where Fig.~\ref{fig6}.\textbf{a} and Fig.~\ref{fig6}.\textbf{b} correspond to the results for 100-ROI and 400-ROI systems, respectively. It is evident that the simulated dynamics with neuronal noise more closely approximate the actual dynamics measured by fMRI, which can be also supported by the statistical result of KL divergence. In contrast, the dynamics without considering noise tends to converge to zero over time, due to the eigenvalues of the coupling matrix $\mathbf{A}$ being all less than zero, leading to the convergence of the state $\mathbf{x}$.

\subsubsection{Control energy of optimal state approaching control}
Although the results of optimal state approaching control and optimal stochastic tracking control cannot be directly compared, we can quantify their relationships with the intrinsic characteristics of the brain system to explore any correlated features. For optimal state approaching control, we used clustering methods to select four representative discrete brain states as target states for both the 100-ROI and 400-ROI brain systems. We then used the peak states of each individual participant as the original states and pushed them to the four target states. Fig.~\ref{fig7}.\textbf{a} and \textbf{d} provide visualizations of the spatial distributions of the four target states, while Fig.~\ref{fig7}.\textbf{b} and \textbf{e} quantify the average distribution of these states within each resting-state system. It can be observed that the four states exhibit distinct spatial distribution characteristics, and the four states in the 100-ROI and 400-ROI brain systems can be matched to each other. Specifically, state 1 of 100-ROI system corresponds to state 4 of 400-ROI system, state 2 of 100-ROI system corresponds to state 1 of 400-ROI system, state 3 of 100-ROI system corresponds to state 2 of 400-ROI system, and state 4 of 100-ROI system corresponds to state 3 of 400-ROI system. We retained the original order of the clustering results for visualization, which does not substantially affect the results.

Fig.~\ref{fig7}.\textbf{c} and \textbf{f} illustrate the relationship between the energy of optimal state approaching control and other features (tracking energy, average controllability, modal controllability, and the value of state itself) across five different groups. Fig.~\ref{fig7}.\textbf{c} presents the results for the 100-ROI brain network system, showing that the control energy required to approach each target state (1 to 4) is predominantly related to the state itself. Similarly, for the 400-ROI brain network system (Fig.~\ref{fig7}.\textbf{f}), despite identifying more statistically significant correlations between the control energy and three features beyond and state value, most significant correlations are concentrated between the energy of optimal state approaching control and the target state itself. These results indicate that the energy of optimal state approaching control is highly dependent on the target state itself, rather than the intrinsic characteristics of the brain network system.

\section{Discussion}
To provide insights for brain stimulation effects and optimal paradigms, this investigation introduces optimal stochastic tracking control for brain networks, reconsidering factors that were not addressed in previous studies. Building upon this foundation, the investigation explores the spatial distribution of stochastic tracking control energy and its relationship with the intrinsic characteristics of the brain network system, while also making indirect comparisons with the results of optimal state approaching control. Finally, we attempt to determine the minimal cost at which the brain network system can achieve acceptable control effect. This investigation proposes a new optimal brain network control framework that is more aligned with brain stimulation objectives and practical brain network models. The following discussion will summarize our research findings.

\subsection{Target dynamics tracking vs. Target state approaching}
The topic of brain network control has evolved significantly since a representative paper by Gu et al. \cite{gu2015controllability}, progressing from studies on network controllability \cite{gu2015controllability,wilmskoetter2022language} to optimal control of linear dynamics \cite{gu2017optimal,fang2022personalizing,stiso2019white,betzel2024controlling}. Gu et al. proposed that cognitive control can be effectively explained by the network controllability properties of specific brain regions, which facilitate the transition of the brain network from one state to a target cognitive state \cite{gu2015controllability}. This sparked considerable interest in whether research could customize input signals to drive the brain along a `controlled' trajectory and into the desired target state. Given the growing interest in brain stimulation techniques (including TMS \cite{hallett2007transcranial}, direct current stimulation \cite{stagg2011physiological}, chemogenetic applications \cite{roth2016dreadds} and optogenetic stimulation \cite{boyden2005millisecond}), this question is highly relevant with numerous clinical applications. Brain states are typically defined by discrete states at a single time point, usually obtained through functional brain activity measurements such as fMRI \cite{fang2022personalizing,betzel2024controlling} and ECoG \cite{stiso2019white}, and then selected through clustering methods to identify representative brain states. While this strategy is not inappropriate for long-term therapy, it is challenging because an individual can produce many brain states even within a short period, and the representativeness of these states cannot be verified. The greatest advantage of optimal tracking control is that the control target is a dynamic process (i.e., a time series). This approach is not affected by state selection and enables the tracking of dynamic characteristics that state approaching control cannot achieve, which potentially represents an important new direction for future brain network optimal control research.

\subsection{Consideration of neuronal noise}
The consideration of neuronal noise is often overlooked in current studies on optimal brain network control. When discussing neuronal noise, they refer to the intrinsic random fluctuations within neural networks \cite{jacobson2005subthreshold}. Under the linear dynamics assumption in current studies \cite{gu2017optimal,fang2022personalizing,stiso2019white,betzel2024controlling}, the brain network is a linear time-invariant system. When noise and external disturbances are not considered, the dynamics are given by $\mathbf{\dot{x}}(t)=\mathbf{A} \mathbf{x}(t)$, where the brain state changes over time solely in relation to the coupled matrix $\mathbf{A}$. In this scenario, the state of the linear system converges to zero when the eigenvalues of the coupling matrix $\mathbf{A}$ are less than zero, or it diverges to infinity, which is inconsistent with the actual dynamics of brain networks. When $\mathbf{w}(t)$ is considered, it includes the stochastic characteristics of neuronal activity at the microscopic level. A similar modeling approach considering noise has been recently studied by Kamiya et al. \cite{kamiya2023optimal} in combination with optimal state approaching control. Although it still cannot fully replicate the actual dynamics of brain networks, the consideration of neuronal noise is also an important advancement in brain network control problem.

\subsection{Correlation of tracking energy with intrinsic brain network characteristics}
Our results provide a clear correlation between tracking energy and the intrinsic characteristics of the brain network system, which may benefit from the complexity of the control target we considered. Although influenced by neuronal noise, the target is a time series with fixed statistical characteristics. After considering both the system and the noise variance, the tracking results are highly robust. The significant negative correlation between tracking energy and average controllability indicates that nodes which can easily drive the network to nearby states require correspondingly higher tracking energy. In contrast, for the optimal state approaching control strategy, the control energy is highly correlated with the selected state values, and its relationship with the intrinsic characteristics of the brain network system is unclear. The state is a discrete vector that, to some extent, is challenging to represent a type of disease or different health states. Therefore, the tracking energy obtained here may offer higher stability and universality for brain stimulation therapy, as it achieves effective dynamics tracking while considering noise, and the results are correlated with brain network characteristics.


\subsection{Modeling dimension of brain network influences tracking strategy}
Our results lead to another new conclusion: different scales of brain network modeling may result in varying control effects with the same proportion of controlled nodes. Existing studies have proposed various brain parcellation templates at different scales \cite{tzourio2002automated,schaefer2018local,yeo2011organization,desikan2006automated}; however, there is no standard criterion. Our research results provide a rough estimate of the minimum tracking cost: if the cortex is modeled in 100 dimensions, selecting the five nodes with the lowest tracking energy for control can achieve a acceptable control effects for dynamics tracking (the dynamics over 90\% nodes can be improved). However, if the system dimension is 400, the control cost (i.e., the number of control nodes required) would significantly increase. In practical brain stimulation techniques such as TMS \cite{hallett2007transcranial}, stimulation localization can be accurate, but this does not mean that the affected region is small. Current research is beginning to model stimulation in a diffusive manner to simulate the impact of brain stimulation on other uncontrolled brain regions \cite{parkes2023using,betzel2024controlling}. Undoubtedly, the extent of the brain region affected by a single stimulation and its diffusion effect will influence the estimation of control cost. 

\subsection{Future directions}
Firstly, the optimal stochastic tracking control strategy we employed is a traditional method in control theory. The optimal control objective simultaneously considers both the tracking effect and the tracking energy. From our results, it is evident that controlling a single node cannot achieve an acceptable tracking effect. The pinning control approach \cite{chen2007pinning,vega2018trajectory} could be attempted in future research, but the increase in tracking energy when reducing the number of control nodes may pose a significant challenge. Secondly, our modeling focused on the entire cortical network, and we did not discuss deeply the mechanisms of stroke and aphasia. Our primary goal was to provide insights for brain stimulation interventions. Future research on post-stroke aphasia could narrow the scope of investigation to consider specific language-related brain regions typically involved such as inferior frontal gyrus (IFG) \cite{fridriksson2016revealing,fridriksson2018anatomy}. Lastly, we believe that other conditions \cite{braun2019brain,stiso2019white,parkes2021network,mahadevan2023alprazolam} requiring short-term or long-term brain stimulation regulation, and more fMRI or electrophysiological data can also be included in the research, as the methodology of this study is not limited to stroke rehabilitation.

\section{Methods}\label{sec11}
\subsection{Data}
\subsubsection{Participants and fMRI acquisition}
Using fMRI data from stroke and post-stroke aphasia cases as examples, we attempted to drive the disease dynamics towards healthy dynamics. We here used two open source fMRI datasets (Aphasia Recovery Cohort (ARC) Dataset, and Resting-state for Older Adults Dataset), which are described below.

\textbf{Dataset 1:} Resting-state fMRI data were obtained from the University of South Carolina and the Medical University of South Carolina \cite{gibson2024aphasia}. The Institutional Review Board (IRB) at the University of South Carolina determined that the study was exempt from further ethical review. The final dataset comprised 148 post-stroke aphasia (including four aphasia types: Anomic, Broca, Conduction, Global) patients and 26 normal stroke patients, after excluding those with head motion greater than 3mm. Participants were instructed to remain still, keep their eyes open, and stay awake during the scan. All fMRI images were collected using a Siemens Trio 3.0T scanner with a 16-channel head coil. For each participant, the fMRI scan closest to the time of stroke onset was selected for further analysis.  

\textbf{Dataset 2:} For finding target states/dynamics for patients, a target group of 28 healthy old participants underwent whole-brain imaging at the Gateway MRI Center at UNCG, utilizing a Siemens 3.0T Tim Trio MRI scanner with a 16-channel head coil, was selected for analysis. The study received approval from the IRB at the University of North Carolina at Greensboro \cite{wahlheim2022intrinsic}. Resting-state fMRI data were acquired following anatomical scans, with 300 measurements collected during the 10-minute scanning sessions. Participants were instructed to remain still, awake, and with their eyes open throughout the procedure. Functional imaging was performed using echo-planar imaging sequences sensitive to blood oxygen level-dependent (BOLD) contrast, with a slice thickness of 4.0 mm, 32 slices, a TE of 30 ms, a TR of 2000 ms, and a flip angle of 70°. The scans covered the entire cerebral cortex and portions of the cerebellum. The primary objective of this dataset was to explore how connectome-based modeling of intrinsic functional connectivity within the default mode network could predict memory discrimination differences between young and older adults. Statistical information of participants for population number, age, and Western Aphasia Battery (WAB) score is presented in Table~\ref{tab1}.

\begin{table}[htbp]
\caption{Participants' information of datasets}\label{tab1}%
\begin{tabular}{@{}llll@{}}
\toprule
Group & Number of population & Age ± (std) & WAB Score ± (std)\\
\midrule
Anomic &   46   & 62 ± 11.1  & 87 ± 6.2   \\
Broca  &   65   & 62 ± 10.7  & 44 ± 17.4  \\
Conduction & 23 & 61 ± 11.7  & 63 ± 15.2  \\
Global &  14   & 62 ± 9.0   & 17 ± 7.7 \\
Normal stroke & 26 & 62 ± 11.9  & 98 ± 1.9 \\
Healthy & 36 & 69.82 ± 5.6   &  Not applicable \\
\botrule
\end{tabular}
\end{table}

\subsubsection{fMRI preprocessing}
For the fMRI raw scan data in the above two datasets, we performed preprocessing operations. The first ten volumes of each scan were excluded to ensure magnetic stability and to optimize the generation of a consistent BOLD activity signal. The preprocessing pipeline was implemented utilizing the CONN toolbox \cite{whitfield2012conn}. Subsequently, the fMRI data underwent the following procedures:

1. Correction for slice timing;

2. Realignment to correct for head motion;

3. Normalization to a 3 mm MNI space using an EPI template provided by the CONN software package;

4. Application of linear temporal detrending to remove the linear drift of the BOLD signal;

5. Covariate regression, including the white matter signal, cerebrospinal fluid (CSF) signal, and head motion signal, utilising the Fristion-24 Parameters strategy;

6. Temporal filtering with a bandwidth of 0.01-0.1 Hz.

We utilized the 100-ROI and 400-ROI parcellation templates from Schaefer2018 parcellation \cite{schaefer2018local} at different scales to parcel the cortex, and for further estimation of brain network systems. We employed a regression-out approach (one of the most used harmonization techniques \cite{gallo2023functional,wang2023comprehensive}) for preprocessed data to eliminate any potential factors that could influence the results, including sites, equipments, acquisition parameters, and the age and gender of individuals.

\subsection{Optimal stochastic tracking control for brain network dynamics}
\subsubsection{Brain network dynamic models}\label{sec2.2}
NCT requires the establishment of brain dynamic model, along with equations describing the connectivity between brain regions. In this context, we employ the stochastic linear time-invariant model:

\begin{equation}
\mathbf{\dot{x}}(t)=\mathbf{A} \mathbf{x}(t) + \mathbf{w}(t),
\label{eq1}
\end{equation}
where $ \mathbf{x}(t) $ is an $ N \times 1 $ vector that represents the brain state (i.e. the BOLD time series for fMRI measurement) at time $t$, and $ N $ is the number of brain regions. In the network adjacency matrix (or coupled matrix) $ \mathbf{A} $, each $ a_{ij} $-th element gives the quantitative anisotropy between region $ i $ and region $ j $. $\mathbf{u}(t)$ is the input which is constant in time.

The coupled matrix $ \mathbf{A} $ in Eq.~\ref{eq1} determines the brain network dynamics, which  
was described by a multivariate Ornstein-Uhlenbeck process \cite{timme2014revealing,gilson2016estimation,gilson2020model}:
\begin{equation}
{\dot{x}}_i(t) = -\frac{{x}_i(t)}{\tau_\mathbf{x}} + \sum_{j \neq i} \mathbf{C}_{ji} {x}_j(t) + {w}_i(t).
\label{eq2}
\end{equation}
where the time constant $\tau_\mathbf{x}$ abstracts the intrinsic dynamics of each ROI, each activity ${x}_i(t)$ of node $i$ decays exponentially according to the time constant $\tau_\mathbf{x}$ and evolves depending on the activity of other populations.
The network effective connectivity (EC) between different ROIs, embodied by the matrix $\mathbf{C}$, whose topological skeleton is determined by structural data. The fluctuating inputs ${w}_i(t)$ are independent and correspond to a diagonal variance matrix $\mathbf{\Sigma}$. In matrix notation, it is:
\begin{equation}
\mathbf{\dot{x}}(t) = \mathbf{J}\mathbf{{x}}(t) + \mathbf{w}(t).
\label{eq3}
\end{equation}

All ${x}_i(t)$ have zero mean, of which the spatiotemporal covariances are denoted by ${Q}_\tau^{ij}$, where $\tau \in \{0, 1\}$ is the time lag, 
and they satisfy the following consistency equations:
\begin{equation}
\mathbf{J}^\top \mathbf{Q}_0 + \mathbf{Q}_0 \mathbf{J} = -\mathbf{\Sigma},
\label{eq4}
\end{equation}
\begin{equation}
\mathbf{Q}_1 = \mathbf{Q}_0 e^\mathbf{J},
\label{eq5}
\end{equation}
where $e$ denotes the matrix exponential, superscript $\top$ denotes the matrix transpose. $\mathbf{J}$ is the Jacobian of the dynamical system, can be considered as a appropriate estimation for $ \mathbf{A} $, which depends on the time constant $\tau_x$ and the network EC:
\begin{equation}
{J}_{ji} = -\frac{\delta_{ji}}{\tau_x} + {C}_{ji},
\label{eq6}
\end{equation}
where $\delta_{ji}$ is the Kronecker delta.

\subsubsection{Parameter estimation of network dynamic model using gradient descent optimization}
Most studies treat the networks extracted from structural imaging as the $\mathbf{A}$ matrix in network control, however, it may not necessarily correspond to the measured brain network dynamics. For obtaining a fine estimation of $\mathbf{J}$ in Eq.~\ref{eq6} for brain network dynamics, i.e. the estimation of $ \mathbf{A} $, we tune the above model in Sec.~\ref{sec2.2} to make its covariance matrices $\mathbf{Q}_0$ and $\mathbf{Q}_1$ reproduce the empirical 
$\mathbf{\hat{Q}}_0$ and $\mathbf{\hat{Q}}_1$ according to previous studies \cite{gilson2016estimation,gilson2020model}, which can be calculated as follows: For a session of $T$ time points, we denote the BOLD time 
series by ${s}_i(t)$ for each region $1 \leq i \leq N$ with time indexed by $1 \leq t \leq T$. 
The mean signal is $\mathbf{\bar{s}}_i = \frac{1}{T} \sum_t {s}_i(t)$ for all $i$. Following previous studies \cite{gilson2016estimation,gilson2020model}, 
two BOLD covariance matrices without and with time lag are:
\begin{equation}
{\hat{Q}}_0^{ij} = \frac{1}{T-2} \sum_{1 \leq t \leq T-1} ({s}_i(t) - {\bar{s}}_i)({s}_j(t) - {\bar{s}}_j),
\label{eq7}
\end{equation}
\begin{equation}
{\hat{Q}}_1^{ij} = \frac{1}{T-2} \sum_{1 \leq t \leq T-1} ({s}_i(t) - {\bar{s}}_i)({s}_j(t+1) - {\bar{s}}_j).
\label{eq8}
\end{equation}

Pearson correlation ${K_{ij}}$ can be calculated from the covariances in Eq.~\ref{eq7} and~\ref{eq8}:
\begin{equation}
{K_{ij}} = \frac{{\hat{Q}}_0^{ij}}{\sqrt{{\hat{Q}}_0^{ii} {\hat{Q}}_0^{jj}}}.
\label{eq9}
\end{equation}

The iterative optimization procedure for $\mathbf{C}$ in Eq.~\ref{eq6} is related to a concept 
of natural gradient descent that accounts for the non-linearity of the mapping between $\mathbf{J}$ and the 
matrix pair $\mathbf{Q}_0$ and $\mathbf{Q}_1$:
\begin{equation}
\tau_{ac} = - \frac{\sum_{i} \log({\hat{Q}}_0^{ii}) - \sum_{i} \log({\hat{Q}}_1^{ii})}{N \cdot \tau_{TR}},
\label{eq10}
\end{equation}
where $\tau_{TR}$ is the time resolution (TR) value of fMRI in seconds.

The difference matrices $\Delta \mathbf{Q}_0 = \mathbf{\hat{Q}}_0 - \mathbf{Q}_0$ and $\Delta \mathbf{Q}_1 = \mathbf{\hat{Q}}_1 - \mathbf{Q}_1$ determine the model error:
\begin{equation}
E = \frac{1}{2} \frac{\|\Delta \mathbf{Q}_0\|^2}{\|\mathbf{\hat{Q}}_0\|^2} + \frac{1}{2} \frac{\|\Delta \mathbf{Q}_1\|^2}{\|\mathbf{\hat{Q}}_1\|^2}.
\label{eq11}
\end{equation}

The Jacobian update is applied to decrease the model error $E$ at each optimization step:
\begin{equation}
\Delta \mathbf{J} = \mathbf{Q}_0^{-1}[-\Delta \mathbf{Q}_0 + \Delta \mathbf{Q}_1 e^\mathbf{J}].
\label{eq12}
\end{equation}

A update that is more robust to empirical noise is:
\begin{equation}
\Delta \mathbf{J} = \mathbf{Q}_0^{-1} \Delta \mathbf{Q}_0 + \Delta \mathbf{Q}_0 \mathbf{Q}_0^{-1} + \mathbf{Q}_1^{-1} \Delta \mathbf{Q}_1 + \Delta \mathbf{Q}_1 \mathbf{Q}_1^{-1}.
\label{eq13}
\end{equation}

Optimal state approaching control requires the coupled matrix $\mathbf{A}$ to be symmetric. To ensure the symmetry of the coupled matrix, we modified the Jacobian update $\Delta \mathbf{J}$ as follows:

\begin{equation}
\Delta \mathbf{J} = \mathbf{Q}_0^{-1} \Delta \mathbf{Q}_0 + \Delta \mathbf{Q}_0 \mathbf{Q}_0^{-1} 
+ 1/2[(\mathbf{Q}_1^{-1} \Delta \mathbf{Q}_1)^\top+\mathbf{Q}_1^{-1} \Delta \mathbf{Q}_1] + 1/2[(\Delta \mathbf{Q}_1 \mathbf{Q}_1^{-1})^\top+\Delta \mathbf{Q}_1 \mathbf{Q}_1^{-1}].
\label{m_eq13}
\end{equation}

From the Jacobian update $\Delta \mathbf{J}$, the connectivity update was obtained:
\begin{equation}
\Delta {C}_{ij} = \eta_\mathbf{C} \Delta {J}_{ij}.
\label{eq14}
\end{equation}

Non-negativity is imposed on the EC values during the optimization. To take properly the effect of cross-correlated inputs into account, the $\mathbf{\Sigma}$ update is:
\begin{equation}
\Delta \mathbf{\Sigma} = -\eta_\mathbf{\Sigma} (\mathbf{J}^\top \Delta \mathbf{Q}_0 + \Delta \mathbf{Q}_0 \mathbf{J}).
\label{eq15}
\end{equation}

Finally, the time constant $\tau_x$ can also be tuned as:
\begin{equation}
\Delta \tau_x = \eta_\tau \left( \tau_{ac} + \frac{1}{\lambda_{\text{max}}} \right),
\label{eq16}
\end{equation}
where $\lambda_{\text{max}}$ is the maximum negative real part of the eigenvalues of $\mathbf{J}$. Repeating the parameter updates, the best fit of matrix $\mathbf{J}$ corresponds to the minimum model error $E$.

\subsubsection{Optimal stochastic tracking control}
Our long-term goal is to use the model described above to predict optimal stimulation parameters, i.e., network control. In the network dynamics control model, we first consider two important factors: one is the stochastic fluctuation in the brain dynamics, which may represent spontaneous activity or noise in brain networks; the other is the optimal control from one network dynamics to a target dynamics, which can be seen as a tracking problem in the control domain, rather than pushing one discrete brain state to another. Given the linear stochastic system state equation like Eq.~\ref{eq1} with input:

\begin{equation}
\mathbf{\dot{x}}(t) = \mathbf{A} \mathbf{x}(t) + \mathbf{B} \mathbf{u}(t) + \mathbf{w}(t).
\end{equation}
where the input matrix $\mathbf{B}$ was defined to allow input dominated by the stimulation region of interest (ROI). The output equation is:

\begin{equation}
\mathbf{z}(t) = \mathbf{D} \mathbf{x}(t),
\end{equation}
where $\mathbf{z}(t)$ is a $N$-dimensional output vector. In brain network systems, the matrix 
$\mathbf{D}$ can be considered as an identity matrix, implying that $\mathbf{z}(t) = \mathbf{x}(t)$.

Let the $N$-dimensional target vector $\mathbf{z}_{r}(t)$ be the output of the target dynamic model under the influence of stochastic noise. The state equation of the target model is known to be:

\begin{equation}
\mathbf{\dot{x}}_{r}(t) = \mathbf{A}_{r} \mathbf{x}_{r}(t) + \mathbf{w}_{r}(t).
\end{equation}
where $\mathbf{w}_{r}(t)$ is stochastic noise with zero mean and variance matrix $\mathbf{Q}_{r}$. The output equation of the target model is:

\begin{equation}
\mathbf{z}_{r}(t) = \mathbf{D} \mathbf{x}_{r}(t),
\end{equation}
where $\mathbf{D}_{r}$ is an identity matrix, implying that $\mathbf{z}_{r}(t) = \mathbf{x}_{r}(t)$.

To take an initial step toward the optimal control goal, we seek to estimate the optimal error and energy required to reach the target brain dynamics, and therefore define the following objective: Determine the linear optimal feedback control $\mathbf{u}^{*}(t)$ such that the system output $\mathbf{z}(t)$ tracks the target dynamics $\mathbf{z}_{r}(t)$ and minimizes the following stochastic performance index:

\begin{equation}
J = E\left\{\frac{1}{2} \int_{t_{0}}^{t_{f}} \left\{ \left[\mathbf{z}(t) - \mathbf{z}_{r}(t)\right]^{\mathrm{T}} \mathbf{Q} \left[\mathbf{z}(t) - \mathbf{z}_{r}(t)\right] + \mathbf{u}^{\mathrm{T}}(t) \mathbf{R} \mathbf{u}(t), \right\} \mathrm{d} t \right\}
\end{equation}
where the final time $t_{f}$ is fixed, and $\mathbf{Q}$ and $\mathbf{R}$ are symmetric non-negative definite and symmetric positive definite matrices, respectively.

For this stochastic output feedback tracking system problem, if the system $\{\mathbf{A}, \mathbf{D}\}$ is completely observable and the target model $\{\mathbf{A}_{r}, \mathbf{D}_{r}\}$ is completely observable, then the optimal control is:

\begin{equation}
\mathbf{u}^{*}(t) = -\mathbf{K}_{1}(t) \mathbf{x}(t) + \mathbf{K}_{2}(t) \mathbf{x}_{r}(t),
\end{equation}
where

\begin{gather}
\mathbf{K}_{1}(t) = \mathbf{R}^{-1} \mathbf{G}^{\mathrm{T}} \mathbf{P}_{11}(t), \\
\mathbf{K}_{2}(t) = -\mathbf{R}^{-1} \mathbf{G}^{\mathrm{T}} \mathbf{P}_{12}(t),
\end{gather}
where $\mathbf{P}_{11}(t)$ and $\mathbf{P}_{12}(t)$ satisfy the following matrix differential equations and their boundary conditions:

\begin{align}
& -\dot{\mathbf{P}}_{11}(t) = \mathbf{P}_{11}(t) \mathbf{A} + \mathbf{A}^{\mathrm{T}} \mathbf{P}_{11}(t) - \mathbf{P}_{11}(t) \mathbf{B} \mathbf{R}^{-1} \mathbf{B}^{\mathrm{T}} \mathbf{P}_{11}(t) + \mathbf{D}^{\mathrm{T}} \mathbf{Q} \mathbf{D}, \\
& \mathbf{P}_{11}(t_{f}) = \mathbf{0}, \\
& -\dot{\mathbf{P}}_{12}(t) = \mathbf{P}_{12}(t) \mathbf{A}_{r} + \mathbf{A}^{\mathrm{T}} \mathbf{P}_{12}(t) - \mathbf{P}_{11}(t) \mathbf{B} \mathbf{R}^{-1} \mathbf{B}^{\mathrm{T}} \mathbf{P}_{12}(t) - \mathbf{D}^{\mathrm{T}} \mathbf{Q} \mathbf{D}_{r}, \\
& \mathbf{P}_{12}(t_{f}) = \mathbf{0}.
\end{align}

The proof and details can be seen in Supplementary Materials.

Here, we introduce several definitions, where Kullback-Leibler (KL) divergence and optimal control energy are employed to assess control effect, while average controllability and modal controllability are used to quantify the intrinsic controllability of brain network systems.

Due to the involvement of stochastic processes in optimal stochastic tracking control, the control effect cannot be measured using Euclidean distances between discrete brain states. Instead, it should be assessed using the KL divergence to quantify the similarity between the statistic of dynamics before/after control and target dynamics. In the context of measuring the similarity between random variables, the KL divergence is a widely used metric in information theory and statistics, and its definition is given as follows: 
\begin{definition}[Kullback–Leibler (KL) divergence \cite{csiszar1975divergence}]
The KL divergence quantifies the difference between two probability distributions. Given two discrete probability distributions \( P \) and \( Q \) defined on the same probability space, the KL divergence from \( Q \) to \( P \) is defined as:

\begin{equation}
D_{\text{KL}}(P \| Q) = \sum_{x \in \mathcal{X}} P(x) \log \left( \frac{P(x)}{Q(x)} \right),
\end{equation}
where \( \mathcal{X} \) denotes the sample space.

For continuous probability distributions with probability density functions \( p \) and \( q \), the KL divergence is given by:

\begin{equation}
D_{\text{KL}}(P \| Q) = \int_{-\infty}^{\infty} p(x) \log \left( \frac{p(x)}{q(x)} \right) \, dx.
\end{equation}
\end{definition}

The KL divergence is always non-negative and is zero if and only if \( P \) and \( Q \) are identical almost everywhere. However, it is important to note that the KL divergence is not a true metric as it is not symmetric, i.e., \( D_{\text{KL}}(P \| Q) \neq D_{\text{KL}}(Q \| P) \), and it does not satisfy the triangle inequality. For each node in brain networks, the KL divergence from the dynamics before control to the target dynamics, and the KL divergence from the dynamics after control to the target dynamics were estimated and compared. We disregarded the estimation of the control effectiveness for optimal state approaching control (which can be directly estimated using methods such as Euclidean distance), because its results cannot be directly compared with those of optimal stochastic tracking control.

\begin{definition}[Optimal control energy]
Following the optimal control theory, the control energy ${E}_{{u}_i^*}$ of each control node can be estimated:
\begin{equation}
    {E}_{{u}_i^*} = \sum_{t=0}^{T-1}( {u}_i^*(t))^2,
\end{equation}
where $T$ is the time points in the control horizon. 
\end{definition}

In the context of optimal stochastic tracking control, we employed a time step of 1 $s$ and 1000 time points for the simulation and calculation of control energy to approximate the resolution used in fMRI sampling. In optimal state approaching control, we similarly used 1000 time points but with a time step of 0.001 $s$ to define the control horizon, as this optimal control method is not suitable for a broad horizon. Consequently, the magnitudes of control energy obtained from the two methods cannot be directly compared, and only their spatial distributions can be produced to correlation analysis.

It is noteworthy that optimal state approaching control aims to drive one brain state to closely approach another target brain state within a limited short time frame. In contrast, optimal stochastic tracking control aims to track an entire brain dynamics towards another target brain dynamics. This highlights the advantage of optimal stochastic tracking control not only in considering the neuronal noise, but also in achieving its intended goal and aligns with the long-term rehabilitation objectives for conditions such as stroke and aphasia.

We quantified the relationship between control energy and the intrinsic controllability of brain network systems. The two types of controllability \cite{gu2015controllability} include:

\begin{definition}[Average controllability\cite{gu2015controllability}]
    Average controllability measures the ease by which input at that node can steer the system into many easily-reachable states, which can be formulated as follows:
\begin{equation} \mathscr{A}_i(\textbf{A})=\sqrt{trace\left(\textbf{B}_i^\top\left(\int_0^{+\infty}\exp(\textbf{A}t^2+2\textbf{A}^\top t)dt\right)\textbf{B}_i\right)},
\end{equation}
where $\textbf{B}_i$ is a one-dimensional input vector that indicates the $i$-th node is controlled.
\end{definition}

\begin{definition}[Modal controllability\cite{gu2015controllability}]
    Modal controllability reflects the ease of a node to push the brain network system into states that is difficult to reach. It can be defined as:
\begin{equation}
    \mathscr{M}_i(\textbf{A})=\sum_{j=1}^N(1-{\lambda_j}^2(\mathbf{A}))v_{ij}^2,
\end{equation}
where $v_{ij}$ is elements in the eigenvector of $\textbf{A}$, and $\lambda_j$ is the $j$-th eigenvalue.
\end{definition}

\subsection{Related work -- Optimal state approaching control for brain network dynamics}

\subsubsection{Brain state selection}
For optimal stochastic tracking control, the target dynamics is the averaged fMRI time series of healthy participants, and the original dynamics is the fMRI time series of each stroke and aphasia patients. For optimal state approaching control, since the state represents a state of a specific moment in time, we employed a clustering method for state selection, following previous research \cite{betzel2024controlling}.

\textbf{Target brain states of healthy participants:}
To define the brain states of healthy participants as target brain states, we referenced the selection method for brain states described in \cite{betzel2024controlling}. Initially, we detected the peaks of the root mean square (RMS) of whole-brain activity amplitude (N = 100 or 400 regions/nodes defined by Schaefer2018 parcellation). We defined a peak as a frame where the RMS exceeds the values of the preceding and following frames. Subsequently, we aggregated the corresponding patterns from participants and calculated the Lin's concordance among all patterns. We then applied a variant of modularity maximization to cluster the $Npeak \times Npeak$ matrix, which resulted in 10 clusters. Most clusters were relatively small and/or contained peaks from only a few participants. We defined a brain state as the cluster center that represents peak activations from at least 50\% of the participants. Ultimately, we identified four distinct brain states for healthy participants. This approach clusters BOLD time series into recurrent states and is similar to established methods for extracting co-activation patterns (CAPs) from BOLD data.

\textbf{Original brain states of stoke and post-stroke aphasia patients:} To define the brain states of patients as original brain states, and to ensure precision and alignment with personalized treatment applications, we calculated an individual brain state for each stroke and aphasia patient. We selected the peaks of the root mean square (RMS) of whole-brain activity amplitude as the brain state, specifically utilizing the z-scored mean of resting-state fMRI (rs-fMRI) signal amplitudes over time for each brain region \cite{weaver2016directional}.

\subsubsection{Optimal state approaching control}

For the traditional optimal brain network control in which the target is to approach the original state to a targeted state (refer to as optimal state approaching control), given the linear deterministic model:

\begin{equation}
    \mathbf{\dot{x}}(t) = \mathbf{A} \mathbf{x}(t) + \mathbf{B} \mathbf{u}(t),
\end{equation}
the cost function is: 
\begin{equation}
    \min_\mathbf{u} \int_0^T \left( (\mathbf{x}_T - \mathbf{x}(t))^\top \mathbf{S} (\mathbf{x}_T - \mathbf{x}(t)) + \rho \mathbf{u}(t)^\top \mathbf{u}(t) \right) \, dt,
    \label{eq17}
\end{equation}
subject to
\begin{align}
    \mathbf{\dot{x}}(t) &= \mathbf{A} \mathbf{x}(t) + \mathbf{B} \mathbf{u}(t), \\
    \mathbf{x}(0) &= \mathbf{x}_0, \\
    \mathbf{x}(T) &= \mathbf{x}_T.
    \label{eq18}
\end{align}

The optimal input can be estimated as follows:

\begin{align}
    \mathbf{u}_k^* &= -\frac{1}{2\rho} \mathbf{B}^\top \mathbf{p}^*, \\
    \mathbf{\dot{x}}^* &= \mathbf{A} \mathbf{x}^* - \frac{1}{2\rho} \mathbf{B} \mathbf{B}^\top \mathbf{p}^*,
    \label{eq21}
\end{align}
where $\mathbf{u}_k^*$ can be obtained through solving corresponding differential equations (see details in Supplementary Materials).

\section{Conclusion}\label{sec13}
In conclusion, we have extended the existing optimal state approaching control of brain networks to optimal stochastic tracking control, which is more aligned with the objectives of brain stimulation. The results indicate that the energy of stochastic tracking control is significantly negatively correlated with the average controllability of the brain network system, whereas the energy of optimal state approaching control is related to the selected target state. Stochastic tracking control of brain networks represents an important direction for future research in brain network control, guiding stimulation paradigms towards minimal cost and optimal therapeutic effect.

\backmatter

\section*{Data availability}
Raw data of the post-stroke aphasia is publicly available at OpenNeuro (\url{https://openneuro.org/datasets/ds004884/versions/1.0.1}). 
Raw data of the healthy participants is publicly available at OpenNeuro (\url{https://openneuro.org/datasets/ds003871/versions/1.0.2}).

\section*{Code availability}
All processing and analysis code is available upon reasonable request.

\section*{Acknowledgments}
This work was supported by the National Natural Science Foundation of China (82172056), Zhejiang Provincial Natural Science Foundation of China (LR23F010003), National Key Research and Development Program of China (2022ZD0117902), the Key Research and Development Program of Zhejiang  Province (2024C04032), and STU Scientific Research Initiation, China (NTF24004T).

\section*{Author contributions}
K.D. and S.C.: conceptualization, methodology, investigation, result visualization, and writing-original draft; K.D., S.C. and Y.D.: investigation and result visualization; K.D., S.C. and L.Z.: methodology and writing-review and editing; X.L., W.L., and Y.Z.: writing-review and editing; K.D. and Y.S.: supervision and writing-review.

\section*{Conflict of interest}
The authors report no competing interests.





\bibliography{sn-article}


\begin{thebibliography}{69}
\ifx \bisbn   \undefined \def \bisbn  #1{ISBN #1}\fi
\ifx \binits  \undefined \def \binits#1{#1}\fi
\ifx \bauthor  \undefined \def \bauthor#1{#1}\fi
\ifx \batitle  \undefined \def \batitle#1{#1}\fi
\ifx \bjtitle  \undefined \def \bjtitle#1{#1}\fi
\ifx \bvolume  \undefined \def \bvolume#1{\textbf{#1}}\fi
\ifx \byear  \undefined \def \byear#1{#1}\fi
\ifx \bissue  \undefined \def \bissue#1{#1}\fi
\ifx \bfpage  \undefined \def \bfpage#1{#1}\fi
\ifx \blpage  \undefined \def \blpage #1{#1}\fi
\ifx \burl  \undefined \def \burl#1{\textsf{#1}}\fi
\ifx \doiurl  \undefined \def \doiurl#1{\url{https://doi.org/#1}}\fi
\ifx \betal  \undefined \def \betal{\textit{et al.}}\fi
\ifx \binstitute  \undefined \def \binstitute#1{#1}\fi
\ifx \binstitutionaled  \undefined \def \binstitutionaled#1{#1}\fi
\ifx \bctitle  \undefined \def \bctitle#1{#1}\fi
\ifx \beditor  \undefined \def \beditor#1{#1}\fi
\ifx \bpublisher  \undefined \def \bpublisher#1{#1}\fi
\ifx \bbtitle  \undefined \def \bbtitle#1{#1}\fi
\ifx \bedition  \undefined \def \bedition#1{#1}\fi
\ifx \bseriesno  \undefined \def \bseriesno#1{#1}\fi
\ifx \blocation  \undefined \def \blocation#1{#1}\fi
\ifx \bsertitle  \undefined \def \bsertitle#1{#1}\fi
\ifx \bsnm \undefined \def \bsnm#1{#1}\fi
\ifx \bsuffix \undefined \def \bsuffix#1{#1}\fi
\ifx \bparticle \undefined \def \bparticle#1{#1}\fi
\ifx \barticle \undefined \def \barticle#1{#1}\fi
\bibcommenthead
\ifx \bconfdate \undefined \def \bconfdate #1{#1}\fi
\ifx \botherref \undefined \def \botherref #1{#1}\fi
\ifx \url \undefined \def \url#1{\textsf{#1}}\fi
\ifx \bchapter \undefined \def \bchapter#1{#1}\fi
\ifx \bbook \undefined \def \bbook#1{#1}\fi
\ifx \bcomment \undefined \def \bcomment#1{#1}\fi
\ifx \oauthor \undefined \def \oauthor#1{#1}\fi
\ifx \citeauthoryear \undefined \def \citeauthoryear#1{#1}\fi
\ifx \endbibitem  \undefined \def \endbibitem {}\fi
\ifx \bconflocation  \undefined \def \bconflocation#1{#1}\fi
\ifx \arxivurl  \undefined \def \arxivurl#1{\textsf{#1}}\fi
\csname PreBibitemsHook\endcsname

\bibitem[\protect\citeauthoryear{Rubinov and Sporns}{2010}]{rubinov2010complex}
\begin{barticle}
\bauthor{\bsnm{Rubinov}, \binits{M.}},
\bauthor{\bsnm{Sporns}, \binits{O.}}:
\batitle{Complex network measures of brain connectivity: uses and interpretations}.
\bjtitle{Neuroimage}
\bvolume{52}(\bissue{3}),
\bfpage{1059}--\blpage{1069}
(\byear{2010})
\end{barticle}
\endbibitem

\bibitem[\protect\citeauthoryear{Whitfield-Gabrieli and Nieto-Castanon}{2012}]{whitfield2012conn}
\begin{barticle}
\bauthor{\bsnm{Whitfield-Gabrieli}, \binits{S.}},
\bauthor{\bsnm{Nieto-Castanon}, \binits{A.}}:
\batitle{Conn: a functional connectivity toolbox for correlated and anticorrelated brain networks}.
\bjtitle{Brain connectivity}
\bvolume{2}(\bissue{3}),
\bfpage{125}--\blpage{141}
(\byear{2012})
\end{barticle}
\endbibitem

\bibitem[\protect\citeauthoryear{Wang et~al.}{2015}]{wang2015gretna}
\begin{barticle}
\bauthor{\bsnm{Wang}, \binits{J.}},
\bauthor{\bsnm{Wang}, \binits{X.}},
\bauthor{\bsnm{Xia}, \binits{M.}},
\bauthor{\bsnm{Liao}, \binits{X.}},
\bauthor{\bsnm{Evans}, \binits{A.}},
\bauthor{\bsnm{He}, \binits{Y.}}:
\batitle{Gretna: a graph theoretical network analysis toolbox for imaging connectomics}.
\bjtitle{Frontiers in human neuroscience}
\bvolume{9},
\bfpage{386}
(\byear{2015})
\end{barticle}
\endbibitem

\bibitem[\protect\citeauthoryear{Baggio et~al.}{2014}]{baggio2014functional}
\begin{barticle}
\bauthor{\bsnm{Baggio}, \binits{H.-C.}},
\bauthor{\bsnm{Sala-Llonch}, \binits{R.}},
\bauthor{\bsnm{Segura}, \binits{B.}},
\bauthor{\bsnm{Marti}, \binits{M.-J.}},
\bauthor{\bsnm{Valldeoriola}, \binits{F.}},
\bauthor{\bsnm{Compta}, \binits{Y.}},
\bauthor{\bsnm{Tolosa}, \binits{E.}},
\bauthor{\bsnm{Junqu{\'e}}, \binits{C.}}:
\batitle{Functional brain networks and cognitive deficits in parkinson's disease}.
\bjtitle{Human brain mapping}
\bvolume{35}(\bissue{9}),
\bfpage{4620}--\blpage{4634}
(\byear{2014})
\end{barticle}
\endbibitem

\bibitem[\protect\citeauthoryear{Kim et~al.}{2017}]{kim2017abnormal}
\begin{barticle}
\bauthor{\bsnm{Kim}, \binits{J.}},
\bauthor{\bsnm{Criaud}, \binits{M.}},
\bauthor{\bsnm{Cho}, \binits{S.S.}},
\bauthor{\bsnm{D{\'\i}ez-Cirarda}, \binits{M.}},
\bauthor{\bsnm{Mihaescu}, \binits{A.}},
\bauthor{\bsnm{Coakeley}, \binits{S.}},
\bauthor{\bsnm{Ghadery}, \binits{C.}},
\bauthor{\bsnm{Valli}, \binits{M.}},
\bauthor{\bsnm{Jacobs}, \binits{M.F.}},
\bauthor{\bsnm{Houle}, \binits{S.}}, \betal:
\batitle{Abnormal intrinsic brain functional network dynamics in parkinson’s disease}.
\bjtitle{Brain}
\bvolume{140}(\bissue{11}),
\bfpage{2955}--\blpage{2967}
(\byear{2017})
\end{barticle}
\endbibitem

\bibitem[\protect\citeauthoryear{Dennis and Thompson}{2014}]{dennis2014functional}
\begin{barticle}
\bauthor{\bsnm{Dennis}, \binits{E.L.}},
\bauthor{\bsnm{Thompson}, \binits{P.M.}}:
\batitle{Functional brain connectivity using fmri in aging and alzheimer’s disease}.
\bjtitle{Neuropsychology review}
\bvolume{24},
\bfpage{49}--\blpage{62}
(\byear{2014})
\end{barticle}
\endbibitem

\bibitem[\protect\citeauthoryear{DelEtoile and Adeli}{2017}]{deletoile2017graph}
\begin{barticle}
\bauthor{\bsnm{DelEtoile}, \binits{J.}},
\bauthor{\bsnm{Adeli}, \binits{H.}}:
\batitle{Graph theory and brain connectivity in alzheimer’s disease}.
\bjtitle{The Neuroscientist}
\bvolume{23}(\bissue{6}),
\bfpage{616}--\blpage{626}
(\byear{2017})
\end{barticle}
\endbibitem

\bibitem[\protect\citeauthoryear{Ashtiani et~al.}{2018}]{ashtiani2018altered}
\begin{barticle}
\bauthor{\bsnm{Ashtiani}, \binits{S.N.M.}},
\bauthor{\bsnm{Daliri}, \binits{M.R.}},
\bauthor{\bsnm{Behnam}, \binits{H.}},
\bauthor{\bsnm{Hossein-Zadeh}, \binits{G.-A.}},
\bauthor{\bsnm{Mehrpour}, \binits{M.}},
\bauthor{\bsnm{Motamed}, \binits{M.R.}},
\bauthor{\bsnm{Fadaie}, \binits{F.}}:
\batitle{Altered topological properties of brain networks in the early ms patients revealed by cognitive task-related fmri and graph theory}.
\bjtitle{Biomedical Signal Processing and Control}
\bvolume{40},
\bfpage{385}--\blpage{395}
(\byear{2018})
\end{barticle}
\endbibitem

\bibitem[\protect\citeauthoryear{Farahani et~al.}{2019}]{farahani2019application}
\begin{barticle}
\bauthor{\bsnm{Farahani}, \binits{F.V.}},
\bauthor{\bsnm{Karwowski}, \binits{W.}},
\bauthor{\bsnm{Lighthall}, \binits{N.R.}}:
\batitle{Application of graph theory for identifying connectivity patterns in human brain networks: a systematic review}.
\bjtitle{frontiers in Neuroscience}
\bvolume{13},
\bfpage{585}
(\byear{2019})
\end{barticle}
\endbibitem

\bibitem[\protect\citeauthoryear{Agosta et~al.}{2014}]{agosta2014disrupted}
\begin{barticle}
\bauthor{\bsnm{Agosta}, \binits{F.}},
\bauthor{\bsnm{Galantucci}, \binits{S.}},
\bauthor{\bsnm{Valsasina}, \binits{P.}},
\bauthor{\bsnm{Canu}, \binits{E.}},
\bauthor{\bsnm{Meani}, \binits{A.}},
\bauthor{\bsnm{Marcone}, \binits{A.}},
\bauthor{\bsnm{Magnani}, \binits{G.}},
\bauthor{\bsnm{Falini}, \binits{A.}},
\bauthor{\bsnm{Comi}, \binits{G.}},
\bauthor{\bsnm{Filippi}, \binits{M.}}:
\batitle{Disrupted brain connectome in semantic variant of primary progressive aphasia}.
\bjtitle{Neurobiology of Aging}
\bvolume{35}(\bissue{11}),
\bfpage{2646}--\blpage{2655}
(\byear{2014})
\end{barticle}
\endbibitem

\bibitem[\protect\citeauthoryear{Tao et~al.}{2020}]{tao2020different}
\begin{barticle}
\bauthor{\bsnm{Tao}, \binits{Y.}},
\bauthor{\bsnm{Ficek}, \binits{B.}},
\bauthor{\bsnm{Rapp}, \binits{B.}},
\bauthor{\bsnm{Tsapkini}, \binits{K.}}:
\batitle{Different patterns of functional network reorganization across the variants of primary progressive aphasia: a graph-theoretic analysis}.
\bjtitle{Neurobiology of aging}
\bvolume{96},
\bfpage{184}--\blpage{196}
(\byear{2020})
\end{barticle}
\endbibitem

\bibitem[\protect\citeauthoryear{Chen et~al.}{2021}]{chen2021disrupted}
\begin{barticle}
\bauthor{\bsnm{Chen}, \binits{X.}},
\bauthor{\bsnm{Chen}, \binits{L.}},
\bauthor{\bsnm{Zheng}, \binits{S.}},
\bauthor{\bsnm{Wang}, \binits{H.}},
\bauthor{\bsnm{Dai}, \binits{Y.}},
\bauthor{\bsnm{Chen}, \binits{Z.}},
\bauthor{\bsnm{Huang}, \binits{R.}}:
\batitle{Disrupted brain connectivity networks in aphasia revealed by resting-state fmri}.
\bjtitle{Frontiers in Aging Neuroscience}
\bvolume{13},
\bfpage{666301}
(\byear{2021})
\end{barticle}
\endbibitem

\bibitem[\protect\citeauthoryear{Bassett and Bullmore}{2006}]{bassett2006small}
\begin{barticle}
\bauthor{\bsnm{Bassett}, \binits{D.S.}},
\bauthor{\bsnm{Bullmore}, \binits{E.}}:
\batitle{Small-world brain networks}.
\bjtitle{The neuroscientist}
\bvolume{12}(\bissue{6}),
\bfpage{512}--\blpage{523}
(\byear{2006})
\end{barticle}
\endbibitem

\bibitem[\protect\citeauthoryear{Gallos et~al.}{2012}]{gallos2012conundrum}
\begin{barticle}
\bauthor{\bsnm{Gallos}, \binits{L.K.}},
\bauthor{\bsnm{Sigman}, \binits{M.}},
\bauthor{\bsnm{Makse}, \binits{H.A.}}:
\batitle{The conundrum of functional brain networks: small-world efficiency or fractal modularity}.
\bjtitle{Frontiers in physiology}
\bvolume{3},
\bfpage{123}
(\byear{2012})
\end{barticle}
\endbibitem

\bibitem[\protect\citeauthoryear{Zamora-L{\'o}pez et~al.}{2010}]{zamora2010cortical}
\begin{barticle}
\bauthor{\bsnm{Zamora-L{\'o}pez}, \binits{G.}},
\bauthor{\bsnm{Zhou}, \binits{C.}},
\bauthor{\bsnm{Kurths}, \binits{J.}}:
\batitle{Cortical hubs form a module for multisensory integration on top of the hierarchy of cortical networks}.
\bjtitle{Frontiers in neuroinformatics}
\bvolume{4},
\bfpage{613}
(\byear{2010})
\end{barticle}
\endbibitem

\bibitem[\protect\citeauthoryear{Guye et~al.}{2010}]{guye2010graph}
\begin{barticle}
\bauthor{\bsnm{Guye}, \binits{M.}},
\bauthor{\bsnm{Bettus}, \binits{G.}},
\bauthor{\bsnm{Bartolomei}, \binits{F.}},
\bauthor{\bsnm{Cozzone}, \binits{P.J.}}:
\batitle{Graph theoretical analysis of structural and functional connectivity mri in normal and pathological brain networks}.
\bjtitle{Magnetic Resonance Materials in Physics, Biology and Medicine}
\bvolume{23},
\bfpage{409}--\blpage{421}
(\byear{2010})
\end{barticle}
\endbibitem

\bibitem[\protect\citeauthoryear{Dong et~al.}{2024}]{dong2024meso}
\begin{barticle}
\bauthor{\bsnm{Dong}, \binits{K.}},
\bauthor{\bsnm{Zhang}, \binits{L.}},
\bauthor{\bsnm{Zhong}, \binits{Y.}},
\bauthor{\bsnm{Xu}, \binits{T.}},
\bauthor{\bsnm{Zhao}, \binits{Y.}},
\bauthor{\bsnm{Chen}, \binits{S.}},
\bauthor{\bsnm{Mahmoud}, \binits{S.S.}},
\bauthor{\bsnm{Fang}, \binits{Q.}}:
\batitle{Meso-scale reorganization of local--global brain networks under mild sedation of propofol anesthesia}.
\bjtitle{NeuroImage}
\bvolume{297},
\bfpage{120744}
(\byear{2024})
\end{barticle}
\endbibitem

\bibitem[\protect\citeauthoryear{Bullmore and Sporns}{2009}]{bullmore2009complex}
\begin{barticle}
\bauthor{\bsnm{Bullmore}, \binits{E.}},
\bauthor{\bsnm{Sporns}, \binits{O.}}:
\batitle{Complex brain networks: graph theoretical analysis of structural and functional systems}.
\bjtitle{Nature reviews neuroscience}
\bvolume{10}(\bissue{3}),
\bfpage{186}--\blpage{198}
(\byear{2009})
\end{barticle}
\endbibitem

\bibitem[\protect\citeauthoryear{Sporns}{2014}]{sporns2014contributions}
\begin{barticle}
\bauthor{\bsnm{Sporns}, \binits{O.}}:
\batitle{Contributions and challenges for network models in cognitive neuroscience}.
\bjtitle{Nature neuroscience}
\bvolume{17}(\bissue{5}),
\bfpage{652}--\blpage{660}
(\byear{2014})
\end{barticle}
\endbibitem

\bibitem[\protect\citeauthoryear{Van Den~Heuvel and Sporns}{2011}]{van2011rich}
\begin{barticle}
\bauthor{\bsnm{Van Den~Heuvel}, \binits{M.P.}},
\bauthor{\bsnm{Sporns}, \binits{O.}}:
\batitle{Rich-club organization of the human connectome}.
\bjtitle{Journal of Neuroscience}
\bvolume{31}(\bissue{44}),
\bfpage{15775}--\blpage{15786}
(\byear{2011})
\end{barticle}
\endbibitem

\bibitem[\protect\citeauthoryear{Schaefer et~al.}{2018}]{schaefer2018local}
\begin{barticle}
\bauthor{\bsnm{Schaefer}, \binits{A.}},
\bauthor{\bsnm{Kong}, \binits{R.}},
\bauthor{\bsnm{Gordon}, \binits{E.M.}},
\bauthor{\bsnm{Laumann}, \binits{T.O.}},
\bauthor{\bsnm{Zuo}, \binits{X.-N.}},
\bauthor{\bsnm{Holmes}, \binits{A.J.}},
\bauthor{\bsnm{Eickhoff}, \binits{S.B.}},
\bauthor{\bsnm{Yeo}, \binits{B.T.}}:
\batitle{Local-global parcellation of the human cerebral cortex from intrinsic functional connectivity mri}.
\bjtitle{Cerebral cortex}
\bvolume{28}(\bissue{9}),
\bfpage{3095}--\blpage{3114}
(\byear{2018})
\end{barticle}
\endbibitem

\bibitem[\protect\citeauthoryear{Hallett}{2007}]{hallett2007transcranial}
\begin{barticle}
\bauthor{\bsnm{Hallett}, \binits{M.}}:
\batitle{Transcranial magnetic stimulation: a primer}.
\bjtitle{Neuron}
\bvolume{55}(\bissue{2}),
\bfpage{187}--\blpage{199}
(\byear{2007})
\end{barticle}
\endbibitem

\bibitem[\protect\citeauthoryear{Feng et~al.}{2013}]{feng2013review}
\begin{barticle}
\bauthor{\bsnm{Feng}, \binits{W.}},
\bauthor{\bsnm{Bowden}, \binits{M.G.}},
\bauthor{\bsnm{Kautz}, \binits{S.}}:
\batitle{Review of transcranial direct current stimulation in poststroke recovery}.
\bjtitle{Topics in stroke rehabilitation}
\bvolume{20}(\bissue{1}),
\bfpage{68}--\blpage{77}
(\byear{2013})
\end{barticle}
\endbibitem

\bibitem[\protect\citeauthoryear{Polan{\'\i}a et~al.}{2018}]{polania2018studying}
\begin{barticle}
\bauthor{\bsnm{Polan{\'\i}a}, \binits{R.}},
\bauthor{\bsnm{Nitsche}, \binits{M.A.}},
\bauthor{\bsnm{Ruff}, \binits{C.C.}}:
\batitle{Studying and modifying brain function with non-invasive brain stimulation}.
\bjtitle{Nature neuroscience}
\bvolume{21}(\bissue{2}),
\bfpage{174}--\blpage{187}
(\byear{2018})
\end{barticle}
\endbibitem

\bibitem[\protect\citeauthoryear{Greene et~al.}{2023}]{greene2023everyone}
\begin{barticle}
\bauthor{\bsnm{Greene}, \binits{A.S.}},
\bauthor{\bsnm{Horien}, \binits{C.}},
\bauthor{\bsnm{Barson}, \binits{D.}},
\bauthor{\bsnm{Scheinost}, \binits{D.}},
\bauthor{\bsnm{Constable}, \binits{R.T.}}:
\batitle{Why is everyone talking about brain state?}
\bjtitle{Trends in Neurosciences}
\bvolume{46}(\bissue{7}),
\bfpage{508}--\blpage{524}
(\byear{2023})
\end{barticle}
\endbibitem

\bibitem[\protect\citeauthoryear{Karrer et~al.}{2020}]{karrer2020practical}
\begin{barticle}
\bauthor{\bsnm{Karrer}, \binits{T.M.}},
\bauthor{\bsnm{Kim}, \binits{J.Z.}},
\bauthor{\bsnm{Stiso}, \binits{J.}},
\bauthor{\bsnm{Kahn}, \binits{A.E.}},
\bauthor{\bsnm{Pasqualetti}, \binits{F.}},
\bauthor{\bsnm{Habel}, \binits{U.}},
\bauthor{\bsnm{Bassett}, \binits{D.S.}}:
\batitle{A practical guide to methodological considerations in the controllability of structural brain networks}.
\bjtitle{Journal of neural engineering}
\bvolume{17}(\bissue{2}),
\bfpage{026031}
(\byear{2020})
\end{barticle}
\endbibitem

\bibitem[\protect\citeauthoryear{Cornblath et~al.}{2020}]{cornblath2020temporal}
\begin{barticle}
\bauthor{\bsnm{Cornblath}, \binits{E.J.}},
\bauthor{\bsnm{Ashourvan}, \binits{A.}},
\bauthor{\bsnm{Kim}, \binits{J.Z.}},
\bauthor{\bsnm{Betzel}, \binits{R.F.}},
\bauthor{\bsnm{Ciric}, \binits{R.}},
\bauthor{\bsnm{Adebimpe}, \binits{A.}},
\bauthor{\bsnm{Baum}, \binits{G.L.}},
\bauthor{\bsnm{He}, \binits{X.}},
\bauthor{\bsnm{Ruparel}, \binits{K.}},
\bauthor{\bsnm{Moore}, \binits{T.M.}}, \betal:
\batitle{Temporal sequences of brain activity at rest are constrained by white matter structure and modulated by cognitive demands}.
\bjtitle{Communications biology}
\bvolume{3}(\bissue{1}),
\bfpage{261}
(\byear{2020})
\end{barticle}
\endbibitem

\bibitem[\protect\citeauthoryear{Parkes et~al.}{2022}]{parkes2022asymmetric}
\begin{barticle}
\bauthor{\bsnm{Parkes}, \binits{L.}},
\bauthor{\bsnm{Kim}, \binits{J.Z.}},
\bauthor{\bsnm{Stiso}, \binits{J.}},
\bauthor{\bsnm{Calkins}, \binits{M.E.}},
\bauthor{\bsnm{Cieslak}, \binits{M.}},
\bauthor{\bsnm{Gur}, \binits{R.E.}},
\bauthor{\bsnm{Gur}, \binits{R.C.}},
\bauthor{\bsnm{Moore}, \binits{T.M.}},
\bauthor{\bsnm{Ouellet}, \binits{M.}},
\bauthor{\bsnm{Roalf}, \binits{D.R.}}, \betal:
\batitle{Asymmetric signaling across the hierarchy of cytoarchitecture within the human connectome}.
\bjtitle{Science Advances}
\bvolume{8}(\bissue{50}),
\bfpage{2185}
(\byear{2022})
\end{barticle}
\endbibitem

\bibitem[\protect\citeauthoryear{Sun et~al.}{2023}]{sun2023network}
\begin{barticle}
\bauthor{\bsnm{Sun}, \binits{H.}},
\bauthor{\bsnm{Jiang}, \binits{R.}},
\bauthor{\bsnm{Dai}, \binits{W.}},
\bauthor{\bsnm{Dufford}, \binits{A.J.}},
\bauthor{\bsnm{Noble}, \binits{S.}},
\bauthor{\bsnm{Spann}, \binits{M.N.}},
\bauthor{\bsnm{Gu}, \binits{S.}},
\bauthor{\bsnm{Scheinost}, \binits{D.}}:
\batitle{Network controllability of structural connectomes in the neonatal brain}.
\bjtitle{Nature communications}
\bvolume{14}(\bissue{1}),
\bfpage{5820}
(\byear{2023})
\end{barticle}
\endbibitem

\bibitem[\protect\citeauthoryear{Stiso et~al.}{2019}]{stiso2019white}
\begin{barticle}
\bauthor{\bsnm{Stiso}, \binits{J.}},
\bauthor{\bsnm{Khambhati}, \binits{A.N.}},
\bauthor{\bsnm{Menara}, \binits{T.}},
\bauthor{\bsnm{Kahn}, \binits{A.E.}},
\bauthor{\bsnm{Stein}, \binits{J.M.}},
\bauthor{\bsnm{Das}, \binits{S.R.}},
\bauthor{\bsnm{Gorniak}, \binits{R.}},
\bauthor{\bsnm{Tracy}, \binits{J.}},
\bauthor{\bsnm{Litt}, \binits{B.}},
\bauthor{\bsnm{Davis}, \binits{K.A.}}, \betal:
\batitle{White matter network architecture guides direct electrical stimulation through optimal state transitions}.
\bjtitle{Cell reports}
\bvolume{28}(\bissue{10}),
\bfpage{2554}--\blpage{2566}
(\byear{2019})
\end{barticle}
\endbibitem

\bibitem[\protect\citeauthoryear{Cooper et~al.}{2021}]{cooper2021mapping}
\begin{barticle}
\bauthor{\bsnm{Cooper}, \binits{R.A.}},
\bauthor{\bsnm{Kurkela}, \binits{K.A.}},
\bauthor{\bsnm{Davis}, \binits{S.W.}},
\bauthor{\bsnm{Ritchey}, \binits{M.}}:
\batitle{Mapping the organization and dynamics of the posterior medial network during movie watching}.
\bjtitle{NeuroImage}
\bvolume{236},
\bfpage{118075}
(\byear{2021})
\end{barticle}
\endbibitem

\bibitem[\protect\citeauthoryear{Braun et~al.}{2019}]{braun2019brain}
\begin{botherref}
\oauthor{\bsnm{Braun}, \binits{U.}},
\oauthor{\bsnm{Harneit}, \binits{A.}},
\oauthor{\bsnm{Pergola}, \binits{G.}},
\oauthor{\bsnm{Menara}, \binits{T.}},
\oauthor{\bsnm{Schaefer}, \binits{A.}},
\oauthor{\bsnm{Betzel}, \binits{R.F.}},
\oauthor{\bsnm{Zang}, \binits{Z.}},
\oauthor{\bsnm{Schweiger}, \binits{J.I.}},
\oauthor{\bsnm{Schwarz}, \binits{K.}},
\oauthor{\bsnm{Chen}, \binits{J.}}, et al.:
Brain state stability during working memory is explained by network control theory, modulated by dopamine d1/d2 receptor function, and diminished in schizophrenia.
bioRxiv,
679670
(2019)
\end{botherref}
\endbibitem

\bibitem[\protect\citeauthoryear{Parkes et~al.}{2021}]{parkes2021network}
\begin{barticle}
\bauthor{\bsnm{Parkes}, \binits{L.}},
\bauthor{\bsnm{Moore}, \binits{T.M.}},
\bauthor{\bsnm{Calkins}, \binits{M.E.}},
\bauthor{\bsnm{Cieslak}, \binits{M.}},
\bauthor{\bsnm{Roalf}, \binits{D.R.}},
\bauthor{\bsnm{Wolf}, \binits{D.H.}},
\bauthor{\bsnm{Gur}, \binits{R.C.}},
\bauthor{\bsnm{Gur}, \binits{R.E.}},
\bauthor{\bsnm{Satterthwaite}, \binits{T.D.}},
\bauthor{\bsnm{Bassett}, \binits{D.S.}}:
\batitle{Network controllability in transmodal cortex predicts positive psychosis spectrum symptoms}.
\bjtitle{Biological Psychiatry}
\bvolume{90}(\bissue{6}),
\bfpage{409}--\blpage{418}
(\byear{2021})
\end{barticle}
\endbibitem

\bibitem[\protect\citeauthoryear{Mahadevan et~al.}{2023}]{mahadevan2023alprazolam}
\begin{barticle}
\bauthor{\bsnm{Mahadevan}, \binits{A.S.}},
\bauthor{\bsnm{Cornblath}, \binits{E.J.}},
\bauthor{\bsnm{Lydon-Staley}, \binits{D.M.}},
\bauthor{\bsnm{Zhou}, \binits{D.}},
\bauthor{\bsnm{Parkes}, \binits{L.}},
\bauthor{\bsnm{Larsen}, \binits{B.}},
\bauthor{\bsnm{Adebimpe}, \binits{A.}},
\bauthor{\bsnm{Kahn}, \binits{A.E.}},
\bauthor{\bsnm{Gur}, \binits{R.C.}},
\bauthor{\bsnm{Gur}, \binits{R.E.}}, \betal:
\batitle{Alprazolam modulates persistence energy during emotion processing in first-degree relatives of individuals with schizophrenia: a network control study}.
\bjtitle{Molecular psychiatry}
\bvolume{28}(\bissue{8}),
\bfpage{3314}--\blpage{3323}
(\byear{2023})
\end{barticle}
\endbibitem

\bibitem[\protect\citeauthoryear{Singleton et~al.}{2022}]{singleton2022receptor}
\begin{barticle}
\bauthor{\bsnm{Singleton}, \binits{S.P.}},
\bauthor{\bsnm{Luppi}, \binits{A.I.}},
\bauthor{\bsnm{Carhart-Harris}, \binits{R.L.}},
\bauthor{\bsnm{Cruzat}, \binits{J.}},
\bauthor{\bsnm{Roseman}, \binits{L.}},
\bauthor{\bsnm{Nutt}, \binits{D.J.}},
\bauthor{\bsnm{Deco}, \binits{G.}},
\bauthor{\bsnm{Kringelbach}, \binits{M.L.}},
\bauthor{\bsnm{Stamatakis}, \binits{E.A.}},
\bauthor{\bsnm{Kuceyeski}, \binits{A.}}:
\batitle{Receptor-informed network control theory links lsd and psilocybin to a flattening of the brain’s control energy landscape}.
\bjtitle{Nature communications}
\bvolume{13}(\bissue{1}),
\bfpage{5812}
(\byear{2022})
\end{barticle}
\endbibitem

\bibitem[\protect\citeauthoryear{Singleton et~al.}{2023}]{singleton2023time}
\begin{botherref}
\oauthor{\bsnm{Singleton}, \binits{S.P.}},
\oauthor{\bsnm{Timmermann}, \binits{C.}},
\oauthor{\bsnm{Luppi}, \binits{A.I.}},
\oauthor{\bsnm{Eckern{\"a}s}, \binits{E.}},
\oauthor{\bsnm{Roseman}, \binits{L.}},
\oauthor{\bsnm{Carhart-Harris}, \binits{R.L.}},
\oauthor{\bsnm{Kuceyeski}, \binits{A.}}:
Time-resolved network control analysis links reduced control energy under dmt with the serotonin 2a receptor, signal diversity, and subjective experience.
bioRxiv
(2023)
\end{botherref}
\endbibitem

\bibitem[\protect\citeauthoryear{Honey et~al.}{2009}]{honey2009predicting}
\begin{barticle}
\bauthor{\bsnm{Honey}, \binits{C.J.}},
\bauthor{\bsnm{Sporns}, \binits{O.}},
\bauthor{\bsnm{Cammoun}, \binits{L.}},
\bauthor{\bsnm{Gigandet}, \binits{X.}},
\bauthor{\bsnm{Thiran}, \binits{J.-P.}},
\bauthor{\bsnm{Meuli}, \binits{R.}},
\bauthor{\bsnm{Hagmann}, \binits{P.}}:
\batitle{Predicting human resting-state functional connectivity from structural connectivity}.
\bjtitle{Proceedings of the National Academy of Sciences}
\bvolume{106}(\bissue{6}),
\bfpage{2035}--\blpage{2040}
(\byear{2009})
\end{barticle}
\endbibitem

\bibitem[\protect\citeauthoryear{Feigin et~al.}{2021}]{feigin2021global}
\begin{barticle}
\bauthor{\bsnm{Feigin}, \binits{V.L.}},
\bauthor{\bsnm{Stark}, \binits{B.A.}},
\bauthor{\bsnm{Johnson}, \binits{C.O.}},
\bauthor{\bsnm{Roth}, \binits{G.A.}},
\bauthor{\bsnm{Bisignano}, \binits{C.}},
\bauthor{\bsnm{Abady}, \binits{G.G.}},
\bauthor{\bsnm{Abbasifard}, \binits{M.}},
\bauthor{\bsnm{Abbasi-Kangevari}, \binits{M.}},
\bauthor{\bsnm{Abd-Allah}, \binits{F.}},
\bauthor{\bsnm{Abedi}, \binits{V.}}, \betal:
\batitle{Global, regional, and national burden of stroke and its risk factors, 1990--2019: a systematic analysis for the global burden of disease study 2019}.
\bjtitle{The Lancet Neurology}
\bvolume{20}(\bissue{10}),
\bfpage{795}--\blpage{820}
(\byear{2021})
\end{barticle}
\endbibitem

\bibitem[\protect\citeauthoryear{Bernhardt et~al.}{2017}]{bernhardt2017agreed}
\begin{barticle}
\bauthor{\bsnm{Bernhardt}, \binits{J.}},
\bauthor{\bsnm{Hayward}, \binits{K.S.}},
\bauthor{\bsnm{Kwakkel}, \binits{G.}},
\bauthor{\bsnm{Ward}, \binits{N.S.}},
\bauthor{\bsnm{Wolf}, \binits{S.L.}},
\bauthor{\bsnm{Borschmann}, \binits{K.}},
\bauthor{\bsnm{Krakauer}, \binits{J.W.}},
\bauthor{\bsnm{Boyd}, \binits{L.A.}},
\bauthor{\bsnm{Carmichael}, \binits{S.T.}},
\bauthor{\bsnm{Corbett}, \binits{D.}}, \betal:
\batitle{Agreed definitions and a shared vision for new standards in stroke recovery research: the stroke recovery and rehabilitation roundtable taskforce}.
\bjtitle{International Journal of Stroke}
\bvolume{12}(\bissue{5}),
\bfpage{444}--\blpage{450}
(\byear{2017})
\end{barticle}
\endbibitem

\bibitem[\protect\citeauthoryear{Varkanitsa and Kiran}{2024}]{varkanitsa2024insights}
\begin{barticle}
\bauthor{\bsnm{Varkanitsa}, \binits{M.}},
\bauthor{\bsnm{Kiran}, \binits{S.}}:
\batitle{Insights gained over 60 years on factors shaping post-stroke aphasia recovery: A commentary on vignolo (1964)}.
\bjtitle{Cortex}
\bvolume{170},
\bfpage{90}--\blpage{100}
(\byear{2024})
\end{barticle}
\endbibitem

\bibitem[\protect\citeauthoryear{Sheppard and Sebastian}{2021}]{sheppard2021diagnosing}
\begin{barticle}
\bauthor{\bsnm{Sheppard}, \binits{S.M.}},
\bauthor{\bsnm{Sebastian}, \binits{R.}}:
\batitle{Diagnosing and managing post-stroke aphasia}.
\bjtitle{Expert review of neurotherapeutics}
\bvolume{21}(\bissue{2}),
\bfpage{221}--\blpage{234}
(\byear{2021})
\end{barticle}
\endbibitem

\bibitem[\protect\citeauthoryear{Wade et~al.}{1986}]{wade1986aphasia}
\begin{barticle}
\bauthor{\bsnm{Wade}, \binits{D.}},
\bauthor{\bsnm{Hewer}, \binits{R.L.}},
\bauthor{\bsnm{David}, \binits{R.M.}},
\bauthor{\bsnm{Enderby}, \binits{P.M.}}:
\batitle{Aphasia after stroke: natural history and associated deficits.}
\bjtitle{Journal of Neurology, Neurosurgery \& Psychiatry}
\bvolume{49}(\bissue{1}),
\bfpage{11}--\blpage{16}
(\byear{1986})
\end{barticle}
\endbibitem

\bibitem[\protect\citeauthoryear{Gu et~al.}{2015}]{gu2015controllability}
\begin{barticle}
\bauthor{\bsnm{Gu}, \binits{S.}},
\bauthor{\bsnm{Pasqualetti}, \binits{F.}},
\bauthor{\bsnm{Cieslak}, \binits{M.}},
\bauthor{\bsnm{Telesford}, \binits{Q.K.}},
\bauthor{\bsnm{Yu}, \binits{A.B.}},
\bauthor{\bsnm{Kahn}, \binits{A.E.}},
\bauthor{\bsnm{Medaglia}, \binits{J.D.}},
\bauthor{\bsnm{Vettel}, \binits{J.M.}},
\bauthor{\bsnm{Miller}, \binits{M.B.}},
\bauthor{\bsnm{Grafton}, \binits{S.T.}}, \betal:
\batitle{Controllability of structural brain networks}.
\bjtitle{Nature communications}
\bvolume{6}(\bissue{1}),
\bfpage{8414}
(\byear{2015})
\end{barticle}
\endbibitem

\bibitem[\protect\citeauthoryear{Gu et~al.}{2017}]{gu2017optimal}
\begin{barticle}
\bauthor{\bsnm{Gu}, \binits{S.}},
\bauthor{\bsnm{Betzel}, \binits{R.F.}},
\bauthor{\bsnm{Mattar}, \binits{M.G.}},
\bauthor{\bsnm{Cieslak}, \binits{M.}},
\bauthor{\bsnm{Delio}, \binits{P.R.}},
\bauthor{\bsnm{Grafton}, \binits{S.T.}},
\bauthor{\bsnm{Pasqualetti}, \binits{F.}},
\bauthor{\bsnm{Bassett}, \binits{D.S.}}:
\batitle{Optimal trajectories of brain state transitions}.
\bjtitle{NeuroImage}
\bvolume{148},
\bfpage{305}--\blpage{317}
(\byear{2017})
\end{barticle}
\endbibitem

\bibitem[\protect\citeauthoryear{Wilmskoetter et~al.}{2022}]{wilmskoetter2022language}
\begin{barticle}
\bauthor{\bsnm{Wilmskoetter}, \binits{J.}},
\bauthor{\bsnm{He}, \binits{X.}},
\bauthor{\bsnm{Caciagli}, \binits{L.}},
\bauthor{\bsnm{Jensen}, \binits{J.H.}},
\bauthor{\bsnm{Marebwa}, \binits{B.}},
\bauthor{\bsnm{Davis}, \binits{K.A.}},
\bauthor{\bsnm{Fridriksson}, \binits{J.}},
\bauthor{\bsnm{Basilakos}, \binits{A.}},
\bauthor{\bsnm{Johnson}, \binits{L.P.}},
\bauthor{\bsnm{Rorden}, \binits{C.}}, \betal:
\batitle{Language recovery after brain injury: a structural network control theory study}.
\bjtitle{Journal of Neuroscience}
\bvolume{42}(\bissue{4}),
\bfpage{657}--\blpage{669}
(\byear{2022})
\end{barticle}
\endbibitem

\bibitem[\protect\citeauthoryear{Fang et~al.}{2022}]{fang2022personalizing}
\begin{barticle}
\bauthor{\bsnm{Fang}, \binits{F.}},
\bauthor{\bsnm{Godlewska}, \binits{B.}},
\bauthor{\bsnm{Cho}, \binits{R.Y.}},
\bauthor{\bsnm{Savitz}, \binits{S.I.}},
\bauthor{\bsnm{Selvaraj}, \binits{S.}},
\bauthor{\bsnm{Zhang}, \binits{Y.}}:
\batitle{Personalizing repetitive transcranial magnetic stimulation for precision depression treatment based on functional brain network controllability and optimal control analysis}.
\bjtitle{NeuroImage}
\bvolume{260},
\bfpage{119465}
(\byear{2022})
\end{barticle}
\endbibitem

\bibitem[\protect\citeauthoryear{Betzel et~al.}{2024}]{betzel2024controlling}
\begin{botherref}
\oauthor{\bsnm{Betzel}, \binits{R.F.}},
\oauthor{\bsnm{Puxeddu}, \binits{M.G.}},
\oauthor{\bsnm{Seguin}, \binits{C.}},
\oauthor{\bsnm{Bazinet}, \binits{V.}},
\oauthor{\bsnm{Luppi}, \binits{A.}},
\oauthor{\bsnm{Podschun}, \binits{A.}},
\oauthor{\bsnm{Singleton}, \binits{S.P.}},
\oauthor{\bsnm{Faskowitz}, \binits{J.}},
\oauthor{\bsnm{Parakkattu}, \binits{V.}},
\oauthor{\bsnm{Misic}, \binits{B.}}, et al.:
Controlling the human connectome with spatially diffuse input signals.
bioRxiv,
2024--02
(2024)
\end{botherref}
\endbibitem

\bibitem[\protect\citeauthoryear{Stagg and Nitsche}{2011}]{stagg2011physiological}
\begin{barticle}
\bauthor{\bsnm{Stagg}, \binits{C.J.}},
\bauthor{\bsnm{Nitsche}, \binits{M.A.}}:
\batitle{Physiological basis of transcranial direct current stimulation}.
\bjtitle{The Neuroscientist}
\bvolume{17}(\bissue{1}),
\bfpage{37}--\blpage{53}
(\byear{2011})
\end{barticle}
\endbibitem

\bibitem[\protect\citeauthoryear{Roth}{2016}]{roth2016dreadds}
\begin{barticle}
\bauthor{\bsnm{Roth}, \binits{B.L.}}:
\batitle{Dreadds for neuroscientists}.
\bjtitle{Neuron}
\bvolume{89}(\bissue{4}),
\bfpage{683}--\blpage{694}
(\byear{2016})
\end{barticle}
\endbibitem

\bibitem[\protect\citeauthoryear{Boyden et~al.}{2005}]{boyden2005millisecond}
\begin{barticle}
\bauthor{\bsnm{Boyden}, \binits{E.S.}},
\bauthor{\bsnm{Zhang}, \binits{F.}},
\bauthor{\bsnm{Bamberg}, \binits{E.}},
\bauthor{\bsnm{Nagel}, \binits{G.}},
\bauthor{\bsnm{Deisseroth}, \binits{K.}}:
\batitle{Millisecond-timescale, genetically targeted optical control of neural activity}.
\bjtitle{Nature neuroscience}
\bvolume{8}(\bissue{9}),
\bfpage{1263}--\blpage{1268}
(\byear{2005})
\end{barticle}
\endbibitem

\bibitem[\protect\citeauthoryear{Jacobson et~al.}{2005}]{jacobson2005subthreshold}
\begin{barticle}
\bauthor{\bsnm{Jacobson}, \binits{G.A.}},
\bauthor{\bsnm{Diba}, \binits{K.}},
\bauthor{\bsnm{Yaron-Jakoubovitch}, \binits{A.}},
\bauthor{\bsnm{Oz}, \binits{Y.}},
\bauthor{\bsnm{Koch}, \binits{C.}},
\bauthor{\bsnm{Segev}, \binits{I.}},
\bauthor{\bsnm{Yarom}, \binits{Y.}}:
\batitle{Subthreshold voltage noise of rat neocortical pyramidal neurones}.
\bjtitle{The Journal of physiology}
\bvolume{564}(\bissue{1}),
\bfpage{145}--\blpage{160}
(\byear{2005})
\end{barticle}
\endbibitem

\bibitem[\protect\citeauthoryear{Kamiya et~al.}{2023}]{kamiya2023optimal}
\begin{barticle}
\bauthor{\bsnm{Kamiya}, \binits{S.}},
\bauthor{\bsnm{Kawakita}, \binits{G.}},
\bauthor{\bsnm{Sasai}, \binits{S.}},
\bauthor{\bsnm{Kitazono}, \binits{J.}},
\bauthor{\bsnm{Oizumi}, \binits{M.}}:
\batitle{Optimal control costs of brain state transitions in linear stochastic systems}.
\bjtitle{Journal of Neuroscience}
\bvolume{43}(\bissue{2}),
\bfpage{270}--\blpage{281}
(\byear{2023})
\end{barticle}
\endbibitem

\bibitem[\protect\citeauthoryear{Tzourio-Mazoyer et~al.}{2002}]{tzourio2002automated}
\begin{barticle}
\bauthor{\bsnm{Tzourio-Mazoyer}, \binits{N.}},
\bauthor{\bsnm{Landeau}, \binits{B.}},
\bauthor{\bsnm{Papathanassiou}, \binits{D.}},
\bauthor{\bsnm{Crivello}, \binits{F.}},
\bauthor{\bsnm{Etard}, \binits{O.}},
\bauthor{\bsnm{Delcroix}, \binits{N.}},
\bauthor{\bsnm{Mazoyer}, \binits{B.}},
\bauthor{\bsnm{Joliot}, \binits{M.}}:
\batitle{Automated anatomical labeling of activations in spm using a macroscopic anatomical parcellation of the mni mri single-subject brain}.
\bjtitle{Neuroimage}
\bvolume{15}(\bissue{1}),
\bfpage{273}--\blpage{289}
(\byear{2002})
\end{barticle}
\endbibitem

\bibitem[\protect\citeauthoryear{Yeo et~al.}{2011}]{yeo2011organization}
\begin{botherref}
\oauthor{\bsnm{Yeo}, \binits{B.T.}},
\oauthor{\bsnm{Krienen}, \binits{F.M.}},
\oauthor{\bsnm{Sepulcre}, \binits{J.}},
\oauthor{\bsnm{Sabuncu}, \binits{M.R.}},
\oauthor{\bsnm{Lashkari}, \binits{D.}},
\oauthor{\bsnm{Hollinshead}, \binits{M.}},
\oauthor{\bsnm{Roffman}, \binits{J.L.}},
\oauthor{\bsnm{Smoller}, \binits{J.W.}},
\oauthor{\bsnm{Z{\"o}llei}, \binits{L.}},
\oauthor{\bsnm{Polimeni}, \binits{J.R.}}, et al.:
The organization of the human cerebral cortex estimated by intrinsic functional connectivity.
Journal of neurophysiology
(2011)
\end{botherref}
\endbibitem

\bibitem[\protect\citeauthoryear{Desikan et~al.}{2006}]{desikan2006automated}
\begin{barticle}
\bauthor{\bsnm{Desikan}, \binits{R.S.}},
\bauthor{\bsnm{S{\'e}gonne}, \binits{F.}},
\bauthor{\bsnm{Fischl}, \binits{B.}},
\bauthor{\bsnm{Quinn}, \binits{B.T.}},
\bauthor{\bsnm{Dickerson}, \binits{B.C.}},
\bauthor{\bsnm{Blacker}, \binits{D.}},
\bauthor{\bsnm{Buckner}, \binits{R.L.}},
\bauthor{\bsnm{Dale}, \binits{A.M.}},
\bauthor{\bsnm{Maguire}, \binits{R.P.}},
\bauthor{\bsnm{Hyman}, \binits{B.T.}}, \betal:
\batitle{An automated labeling system for subdividing the human cerebral cortex on mri scans into gyral based regions of interest}.
\bjtitle{Neuroimage}
\bvolume{31}(\bissue{3}),
\bfpage{968}--\blpage{980}
(\byear{2006})
\end{barticle}
\endbibitem

\bibitem[\protect\citeauthoryear{Parkes et~al.}{2023}]{parkes2023using}
\begin{botherref}
\oauthor{\bsnm{Parkes}, \binits{L.}},
\oauthor{\bsnm{Kim}, \binits{J.Z.}},
\oauthor{\bsnm{Stiso}, \binits{J.}},
\oauthor{\bsnm{Brynildsen}, \binits{J.K.}},
\oauthor{\bsnm{Cieslak}, \binits{M.}},
\oauthor{\bsnm{Covitz}, \binits{S.}},
\oauthor{\bsnm{Gur}, \binits{R.E.}},
\oauthor{\bsnm{Gur}, \binits{R.C.}},
\oauthor{\bsnm{Pasqualetti}, \binits{F.}},
\oauthor{\bsnm{Shinohara}, \binits{R.T.}}, et al.:
Using network control theory to study the dynamics of the structural connectome.
bioRxiv
(2023)
\end{botherref}
\endbibitem

\bibitem[\protect\citeauthoryear{Chen et~al.}{2007}]{chen2007pinning}
\begin{barticle}
\bauthor{\bsnm{Chen}, \binits{T.}},
\bauthor{\bsnm{Liu}, \binits{X.}},
\bauthor{\bsnm{Lu}, \binits{W.}}:
\batitle{Pinning complex networks by a single controller}.
\bjtitle{IEEE Transactions on Circuits and Systems I: Regular Papers}
\bvolume{54}(\bissue{6}),
\bfpage{1317}--\blpage{1326}
(\byear{2007})
\end{barticle}
\endbibitem

\bibitem[\protect\citeauthoryear{Vega et~al.}{2018}]{vega2018trajectory}
\begin{barticle}
\bauthor{\bsnm{Vega}, \binits{C.J.}},
\bauthor{\bsnm{Suarez}, \binits{O.J.}},
\bauthor{\bsnm{Sanchez}, \binits{E.N.}},
\bauthor{\bsnm{Chen}, \binits{G.}},
\bauthor{\bsnm{Elvira-Ceja}, \binits{S.}},
\bauthor{\bsnm{Rodriguez-Castellanos}, \binits{D.}}:
\batitle{Trajectory tracking on complex networks via inverse optimal pinning control}.
\bjtitle{IEEE Transactions on Automatic Control}
\bvolume{64}(\bissue{2}),
\bfpage{767}--\blpage{774}
(\byear{2018})
\end{barticle}
\endbibitem

\bibitem[\protect\citeauthoryear{Fridriksson et~al.}{2016}]{fridriksson2016revealing}
\begin{barticle}
\bauthor{\bsnm{Fridriksson}, \binits{J.}},
\bauthor{\bsnm{Yourganov}, \binits{G.}},
\bauthor{\bsnm{Bonilha}, \binits{L.}},
\bauthor{\bsnm{Basilakos}, \binits{A.}},
\bauthor{\bsnm{Den~Ouden}, \binits{D.-B.}},
\bauthor{\bsnm{Rorden}, \binits{C.}}:
\batitle{Revealing the dual streams of speech processing}.
\bjtitle{Proceedings of the National Academy of Sciences}
\bvolume{113}(\bissue{52}),
\bfpage{15108}--\blpage{15113}
(\byear{2016})
\end{barticle}
\endbibitem

\bibitem[\protect\citeauthoryear{Fridriksson et~al.}{2018}]{fridriksson2018anatomy}
\begin{barticle}
\bauthor{\bsnm{Fridriksson}, \binits{J.}},
\bauthor{\bsnm{Ouden}, \binits{D.-B.}},
\bauthor{\bsnm{Hillis}, \binits{A.E.}},
\bauthor{\bsnm{Hickok}, \binits{G.}},
\bauthor{\bsnm{Rorden}, \binits{C.}},
\bauthor{\bsnm{Basilakos}, \binits{A.}},
\bauthor{\bsnm{Yourganov}, \binits{G.}},
\bauthor{\bsnm{Bonilha}, \binits{L.}}:
\batitle{Anatomy of aphasia revisited}.
\bjtitle{Brain}
\bvolume{141}(\bissue{3}),
\bfpage{848}--\blpage{862}
(\byear{2018})
\end{barticle}
\endbibitem

\bibitem[\protect\citeauthoryear{Gibson et~al.}{2024}]{gibson2024aphasia}
\begin{barticle}
\bauthor{\bsnm{Gibson}, \binits{M.}},
\bauthor{\bsnm{Newman-Norlund}, \binits{R.}},
\bauthor{\bsnm{Bonilha}, \binits{L.}},
\bauthor{\bsnm{Fridriksson}, \binits{J.}},
\bauthor{\bsnm{Hickok}, \binits{G.}},
\bauthor{\bsnm{Hillis}, \binits{A.E.}},
\bauthor{\bsnm{Ouden}, \binits{D.-B.}},
\bauthor{\bsnm{Rorden}, \binits{C.}}:
\batitle{The aphasia recovery cohort, an open-source chronic stroke repository}.
\bjtitle{Scientific Data}
\bvolume{11}(\bissue{1}),
\bfpage{981}
(\byear{2024})
\end{barticle}
\endbibitem

\bibitem[\protect\citeauthoryear{Wahlheim et~al.}{2022}]{wahlheim2022intrinsic}
\begin{barticle}
\bauthor{\bsnm{Wahlheim}, \binits{C.N.}},
\bauthor{\bsnm{Christensen}, \binits{A.P.}},
\bauthor{\bsnm{Reagh}, \binits{Z.M.}},
\bauthor{\bsnm{Cassidy}, \binits{B.S.}}:
\batitle{Intrinsic functional connectivity in the default mode network predicts mnemonic discrimination: A connectome-based modeling approach}.
\bjtitle{Hippocampus}
\bvolume{32}(\bissue{1}),
\bfpage{21}--\blpage{37}
(\byear{2022})
\end{barticle}
\endbibitem

\bibitem[\protect\citeauthoryear{Gallo et~al.}{2023}]{gallo2023functional}
\begin{barticle}
\bauthor{\bsnm{Gallo}, \binits{S.}},
\bauthor{\bsnm{El-Gazzar}, \binits{A.}},
\bauthor{\bsnm{Zhutovsky}, \binits{P.}},
\bauthor{\bsnm{Thomas}, \binits{R.M.}},
\bauthor{\bsnm{Javaheripour}, \binits{N.}},
\bauthor{\bsnm{Li}, \binits{M.}},
\bauthor{\bsnm{Bartova}, \binits{L.}},
\bauthor{\bsnm{Bathula}, \binits{D.}},
\bauthor{\bsnm{Dannlowski}, \binits{U.}},
\bauthor{\bsnm{Davey}, \binits{C.}}, \betal:
\batitle{Functional connectivity signatures of major depressive disorder: machine learning analysis of two multicenter neuroimaging studies}.
\bjtitle{Molecular Psychiatry}
\bvolume{28}(\bissue{7}),
\bfpage{3013}--\blpage{3022}
(\byear{2023})
\end{barticle}
\endbibitem

\bibitem[\protect\citeauthoryear{Wang et~al.}{2023}]{wang2023comprehensive}
\begin{barticle}
\bauthor{\bsnm{Wang}, \binits{Y.-W.}},
\bauthor{\bsnm{Chen}, \binits{X.}},
\bauthor{\bsnm{Yan}, \binits{C.-G.}}:
\batitle{Comprehensive evaluation of harmonization on functional brain imaging for multisite data-fusion}.
\bjtitle{NeuroImage}
\bvolume{274},
\bfpage{120089}
(\byear{2023})
\end{barticle}
\endbibitem

\bibitem[\protect\citeauthoryear{Timme and Casadiego}{2014}]{timme2014revealing}
\begin{barticle}
\bauthor{\bsnm{Timme}, \binits{M.}},
\bauthor{\bsnm{Casadiego}, \binits{J.}}:
\batitle{Revealing networks from dynamics: an introduction}.
\bjtitle{Journal of Physics A: Mathematical and Theoretical}
\bvolume{47}(\bissue{34}),
\bfpage{343001}
(\byear{2014})
\end{barticle}
\endbibitem

\bibitem[\protect\citeauthoryear{Gilson et~al.}{2016}]{gilson2016estimation}
\begin{barticle}
\bauthor{\bsnm{Gilson}, \binits{M.}},
\bauthor{\bsnm{Moreno-Bote}, \binits{R.}},
\bauthor{\bsnm{Ponce-Alvarez}, \binits{A.}},
\bauthor{\bsnm{Ritter}, \binits{P.}},
\bauthor{\bsnm{Deco}, \binits{G.}}:
\batitle{Estimation of directed effective connectivity from fmri functional connectivity hints at asymmetries of cortical connectome}.
\bjtitle{PLoS computational biology}
\bvolume{12}(\bissue{3}),
\bfpage{1004762}
(\byear{2016})
\end{barticle}
\endbibitem

\bibitem[\protect\citeauthoryear{Gilson et~al.}{2020}]{gilson2020model}
\begin{barticle}
\bauthor{\bsnm{Gilson}, \binits{M.}},
\bauthor{\bsnm{Zamora-L{\'o}pez}, \binits{G.}},
\bauthor{\bsnm{Pallar{\'e}s}, \binits{V.}},
\bauthor{\bsnm{Adhikari}, \binits{M.H.}},
\bauthor{\bsnm{Senden}, \binits{M.}},
\bauthor{\bsnm{Campo}, \binits{A.T.}},
\bauthor{\bsnm{Mantini}, \binits{D.}},
\bauthor{\bsnm{Corbetta}, \binits{M.}},
\bauthor{\bsnm{Deco}, \binits{G.}},
\bauthor{\bsnm{Insabato}, \binits{A.}}:
\batitle{Model-based whole-brain effective connectivity to study distributed cognition in health and disease}.
\bjtitle{Network Neuroscience}
\bvolume{4}(\bissue{2}),
\bfpage{338}--\blpage{373}
(\byear{2020})
\end{barticle}
\endbibitem

\bibitem[\protect\citeauthoryear{Csisz{\'a}r}{1975}]{csiszar1975divergence}
\begin{botherref}
\oauthor{\bsnm{Csisz{\'a}r}, \binits{I.}}:
I-divergence geometry of probability distributions and minimization problems.
The annals of probability,
146--158
(1975)
\end{botherref}
\endbibitem

\bibitem[\protect\citeauthoryear{Weaver et~al.}{2016}]{weaver2016directional}
\begin{barticle}
\bauthor{\bsnm{Weaver}, \binits{K.E.}},
\bauthor{\bsnm{Wander}, \binits{J.D.}},
\bauthor{\bsnm{Ko}, \binits{A.L.}},
\bauthor{\bsnm{Casimo}, \binits{K.}},
\bauthor{\bsnm{Grabowski}, \binits{T.J.}},
\bauthor{\bsnm{Ojemann}, \binits{J.G.}},
\bauthor{\bsnm{Darvas}, \binits{F.}}:
\batitle{Directional patterns of cross frequency phase and amplitude coupling within the resting state mimic patterns of fmri functional connectivity}.
\bjtitle{Neuroimage}
\bvolume{128},
\bfpage{238}--\blpage{251}
(\byear{2016})
\end{barticle}
\endbibitem

\end{thebibliography}

\end{document}